\documentclass[12pt,preprint]{aastex}

\shorttitle{Mode Identification from Combinations}
\shortauthors{Yeates et al.}
\received{}

\begin{document}

\title{Mode Identification from Combination Frequency Amplitudes in ZZ Ceti Stars}

\author{Celeste M. Yeates\altaffilmark{1}, J. Christopher Clemens\altaffilmark{1}, S. E. Thompson\altaffilmark{2} \& F. Mullally\altaffilmark{3,4}}

\altaffiltext{1}{Department of Physics and Astronomy, University of North 
Carolina, CB \#3255, Chapel Hill, NC 27599; yeates@physics.unc.edu, clemens@physics.unc.edu}
\altaffiltext{2}{Department of Physics, Colorado College, 14 E. Cache La Poudre, Colorado Springs, CO 80903; sthompson@coloradocollege.edu}
\altaffiltext{3}{Department of Astronomy, University of Texas, 1 University Station, C1400, Austin, TX 78712; fergal@astro.as.utexas.edu}
\altaffiltext{4}{McDonald Observatory, Fort Davis, TX 79734}

\begin{abstract} 

The lightcurves of variable DA stars are usually multi-periodic and non-sinusoidal, so that their Fourier transforms show peaks at eigenfrequencies of the pulsation modes and at sums and differences of these frequencies.  These combination frequencies provide extra information about the pulsations, both physical and geometrical, that is lost unless they are analyzed.  Several theories provide a context for this analysis by predicting combination frequency amplitudes.  In these theories, the combination frequencies arise from nonlinear mixing of oscillation modes in the outer layers of the white dwarf, so their analysis cannot yield direct information on the global structure of the star as eigenmodes provide.  However, their sensitivity to mode geometry does make them a useful tool for identifying the spherical degree of the modes that mix to produce them.  In this paper, we analyze data from eight hot, low-amplitude DAV white dwarfs and measure the amplitudes of combination frequencies present.  By comparing these amplitudes to the predictions of the theory of Goldreich and Wu, we have verified that the theory is crudely consistent with the measurements.  We have also investigated to what extent the combination frequencies can be used to measure the spherical degree ($\ell$) of the modes that produce them.  We find that modes with $\ell > 2$ are easily identifiable as high $\ell$ based on their combination frequencies alone.  Distinguishing between $\ell=1$ and 2 is also possible using harmonics.  These results will be useful for conducting seismological analysis of large ensembles of ZZ Ceti stars, such as those being discovered using the Sloan Digital Sky Survey.  Because this method relies only on photometry at optical wavelengths, it can be applied to faint stars using four-meter class telescopes.

\end{abstract}
\keywords{stars: individual(\objectname{GD 66}, \objectname[LP 521-049]{GD 244}, \objectname[RY LMi]{G117-B15A}, \objectname[PY Vul]{G185-32}, \objectname[MY Aps]{L19-2}, \objectname{GD 165}, \objectname[ZZ Cet]{R548}, \objectname[DN Dra]{G226-29}) --- stars: oscillations --- stars: variables: other --- white dwarfs }

\section{INTRODUCTION}
\label{intro}

There are three known classes of pulsating white dwarf stars in three different instability strips: the pulsating PG 1159 stars at about 100,000 K, the DBV (He I spectrum, variable) stars at 25,000 K, and the DAV (H) stars at 12,000 K.  In spite of the differences in temperature and surface composition, the pulsation periods and the appearance of the lightcurves are similar.  The DAV and DBV stars in particular (with periods between 100 and 1000 seconds), have distinctive non-sinusoidal variations at large amplitude, and more linear behavior at small amplitude \citep{mc80}.  As a consequence, the Fourier transforms of DAV and DBV lightcurves generally show power at harmonics and at sum and difference frequencies.  These ``combination frequencies'' are not in general the result of independent pulsation eigenmodes, but rather of frequency mixing between eigenmodes \citep{brick92b,gw99,ik01,bfw95}.  In this paper, we will explore combination frequencies in the small amplitude DAV (ZZ Ceti) white dwarf stars.  The combination frequency peaks are smaller, and therefore harder to detect, than the combination frequencies in large amplitude pulsators like G29-38, but they are more stable, and therefore more likely to yield understandable and repeatable results.  The small amplitude DAVs are also the pulsators in which the origin of the combination frequencies are most uncertain---they might arise from convective effects \citep{brick92b} or from nonlinearities in the radiative flux alone, as proposed by \citet{bfw95}.

The spectroscopic measurements of \citet{gre76,gre82} showed that the ZZ Ceti stars lie within a narrow range of effective temperatures, and \citet{fon82}~ reasoned that most, if not all, DA white dwarfs are variables as they cool through this instability strip.  The low amplitude pulsators we consider in this paper lie at the high temperature end of the instability strip, and have short, stable pulsation periods \citep{wf82,clemens94}.  The variations we observe arise from the temperature changes associated with non-radial gravity-mode pulsations \citep{rkn82}.  At some amplitude, these pulsations will appear non-sinusoidal because of the $T^4$ dependence of the measured flux.  The combination frequencies that we measure in the low amplitude DAV white dwarfs discussed in this paper are larger than those expected from the $T^4$ nonlinearity, and require an additional nonlinear process in the surface layers of the white dwarf.

The first attempt to identify the nonlinear process was \citet{brick83,brick90,brick91a,brick91b,brick92a,brick92b}, who explored the time dependent properties of the surface convection zone.  Using a numerical model of the surface convection zone, \citet{brick92a} calculated the first non-sinusoidal theoretical shapes of ZZ Ceti lightcurves.  In his model, the nearly isentropic surface convection zone adjusts its entropy on short timescales, attenuating and delaying any flux changes that originate at its base.  As the convection zone changes thickness during a single pulsation cycle, the amount of attenuation and delay changes as well, distorting sinusoidal input variations and creating combination frequencies in the Fourier spectrum of the output signal.

\citet{gw99} repeated and expanded Brickhill's work using an analytic approach.  \citet[hereafter Wu]{wu} was able to derive approximate expressions for the size of combination frequencies that depend upon the frequency, amplitude, and spherical harmonic indices of the parent modes, and upon the inclination of the star's pulsation axis to our line of sight.  Her solutions yield physical insight into the problem, and make predictions for individual stars straightforward to calculate.  Wu herself compared her calculations to measured combination frequencies in the DBV GD 358 and the large-amplitude ZZ Ceti, G29-38, finding good correspondence.

Subsequently, \citet{ik01} extended the numerical simulations of lightcurves of \citet{brick92a,brick92b}, showing that for large amplitude pulsations ($\delta P \diagup P > 5 \%$) the numerical models must incorporate the time-dependence of quantities that are held constant in the method of Brickhill (e.g., heat capacities).  In these full time-dependent calculations, the large amplitude variations begin to show maxima in locations different from those described by the low-order spherical harmonics.  However, for the small amplitude variations we consider in this paper, this effect is negligible, and the numerical results of \citet{ik01} are in agreement with \citet{brick92a,brick92b} and Wu.

An entirely different model for explaining combination frequencies was proposed by \citet[hereafter BFW]{bfw95}.  Instead of changes in the convection zone, BFW invoke the nonlinear response of the radiative atmosphere, ignoring the changes to the surface convection zone.  These radiative nonlinearities can be larger than expected from the $T^4$ dependence of flux because of the sensitivity of the H absorption lines to temperature.  \citet{vb00} compared the predictions of this theory to those of Brickhill for the large amplitude pulsator G29-38, and found that the combination frequencies in that star are too large to be explained by the BFW theory.  This does not necessarily invalidate the theory, but suggests that some other mechanism is at work, at least in G29-38.  \citet{vb00} left open the question of low amplitude pulsators, which have much smaller combination frequencies.  In one case at least (G117-B15A), the BFW theory was able to account for the amplitude of the combination frequencies \citep{brass93}.  However, this success relied on an exact match between the spectroscopic temperature of the star and a narrow maximum in the theoretical predictions.  Using more recent spectroscopic temperature estimates for G117-B15A, which differ from the old by only 850 K \citep[see][]{ber04}, the theory underestimates the combination frequency amplitudes by more than an order of magnitude.  In general, even for the low amplitude pulsators, the BFW theory underestimates the sizes of combination frequencies by an order of magnitude or more. 

In this paper we compare the analytic theory of Wu to observations of the hot, low-amplitude ZZ Ceti stars GD 66, GD 244, G117-B15A, G185-32, L19-2, GD 165, R548, and G226-29, to determine how well this theory reproduces the combination frequencies in hot DAV stars.  If they correctly describe and predict the behavior of DAV pulsations, the analytical formulae of Wu will be an important tool for mode identification in ZZ Ceti stars.  Confidently assigning values of the spherical degree ($\ell$) and azimuthal order ($m$) to individual eigenfrequencies has heretofore required time-resolved spectroscopy using either very large optical telescopes or the Hubble Space Telescope \citep{cvw00,rob95}.  As \citet{brick92b} first proposed, and BFW reiterated, a reliable theory for combination frequency amplitudes allows pulsation mode identification based on measurements of combination frequencies alone.  Thus, if we can verify that the predictions of Wu are consistent with observations, then they constitute an uncomplicated method of mode identification that relies only upon broadband photometry rather than spectroscopy.  Moreover, the theory of Goldreich \& Wu (and of Brickhill) implicitly contains a mode driving mechanism different from that originally proposed for the DAV stars.  Verification of the analytical predictions of Wu will support this convective driving mechanism as the source of pulsations in DAV stars.

For this paper, we have used published Fourier spectra, new reductions of archival Whole Earth Telescope data, and original data obtained with the McDonald Observatory 2.1-m Struve telescope to measure combination frequency amplitudes or amplitude limits for eight hot DAV stars.  We have applied our best estimates of the inclination of the pulsation axis to the observer's line of sight and compared the amplitudes of the combination frequencies to the analytical calculations of Wu.  We find that the theory reproduces the relative amplitudes of combination frequencies in these stars very well, but over-predicts their absolute values by a factor of about 1.4.  The calculations of Wu include an adjustable parameterization of the radiative atmosphere which can easily accommodate a factor this large.  When we normalize its value using the star in our sample with the most detected combination frequencies, GD 66, the theory reproduces all the observed ratios of combination frequency to parent mode amplitudes to better than a factor of two.  This is easily sufficient to verify the high $\ell$ identification for modes in G185-32, as established by \citet{thomp04} using time-resolved spectroscopy from Keck and the Hubble Space Telescope.  It is also sufficient in many cases to distinguish between $\ell=1$ and $\ell=2$ by relying on mode harmonics.  Based on these results, we conclude that the theory of Wu, suitably calibrated, can function as a mode identification method for at least the hot ZZ Ceti stars.  This conclusion is important because the follow-up photometry of ZZ Ceti candidates from the Sloan Digital Sky Survey is finding large numbers of these pulsators that will be too faint for practical time-resolved spectroscopic methods \citep[see][]{muk04,mul05}.  Our results suggest that time-series photometry on four-meter class telescopes, augmented with multiplet splitting where available \citep[e.g.,][]{brad05}, will be sufficient to classify modes in these stars.

We begin in \S\ref{theory} by summarizing the analytical expressions of Wu necessary for predicting the amplitudes of combination frequencies.  We also discuss our method for estimating the inclination of the stars' pulsation axes to the observer's line of sight, and show that our result is insensitive to error in this estimate.  In \S\ref{data} we present the data for each of the eight stars individually and compare predictions based on Wu's equations with the observed amplitudes.  In \S\ref{sum} we summarize our results and discuss future application of the technique, emphasizing a prescription for applying the theory of Wu to large samples of ZZ Ceti stars.

\section{THEORETICAL REVIEW}
\label{theory}

In this section, we will summarize Wu's analytic model and explain how we apply her theory, along with an independent estimation of the stellar inclination angle, to predict the amplitudes of combination frequencies.  Wu's models rely upon an attenuation and a delay of the perturbed flux within the convection zone to produce non-sinusoidal photospheric flux variations.  The differential equation describing these effects (Wu) is:

\begin{equation}
\label{diff}
{\left({{\delta F}\over{F}}\right)_b} = {X + \tau_{c_\circ}[1 + (2\beta + \gamma)X]{{dX}\over{dt}}},
\end{equation}

\noindent where $(\delta F \diagup F)_b$ is the assumed sinusoidal flux perturbation at the base of the convection zone.  $X \equiv (\delta F \diagup F)_{ph}$ is the flux variation at the photosphere and is related to the photometric variations we observe.  $\tau_{c_\circ}$ is the time delay introduced by the convection zone.  Physically it represents the timescale over which the convection zone can absorb a flux change (by adjusting its entropy) instead of communicating it to the surface.  In this paper we approximate $\tau_{c_\circ}$ by setting it equal to the longest observed mode period.  This is a lower limit, because modes with periods longer than $\tau_{c_\circ}$ cannot be driven, but the longest observed period might not be quite as large as $\tau_{c_\circ}$.  $\beta$ and $\gamma$ are fixed parameterizations of the radiative region overlying the convection zone, and represent an attenuation of the flux.  The mixing length models of \citet{wg99} yield $\beta \sim 1.2$ and $\gamma \sim -15$ in the temperature range of ZZ Ceti stars (see their Figure 1).  

The solution to equation~\ref{diff} represents the detectable flux variation at the photosphere, and has the assumed form:

\begin{eqnarray}
\label{solution}
{\left({{\delta F}\over{F}}\right)_{ph}} = a_i\cos(\omega_it+\psi_i)+a_{2i}\cos(2\omega_it+\psi_{2i}) +a_j\cos(\omega_jt+\psi_j)+a_{2j}\cos(2\omega_jt+\psi_{2j}) \nonumber \\ +a_{i-j}\cos[(\omega_i-\omega_j)t+\psi_{i-j}] +a_{i+j}\cos[(\omega_i+\omega_j)t+\psi_{i+j}]+...
\end{eqnarray}

\noindent Solving equation~\ref{diff} yields expressions for the amplitude coefficients ($a_{i\pm j}$) and the phases ($\psi_{i\pm j}$) at each combination frequency ($\omega_i\pm \omega_j$).  In this paper we do not consider phases because they are impossible to recover from some of the published data, and difficult to measure in the presence of noise.  Thus we focus on the amplitudes represented by:

\begin{eqnarray}
\label{combamp}
a_{i\pm j} = {{n_{ij}}\over{2}} {{a_i a_j}\over{2}} {{\mid 2\beta+\gamma \mid (\omega_i \pm \omega_j)\tau_{c_\circ}}\over{\sqrt{1+[(\omega_i \pm \omega_j)\tau_{c_\circ}]^2}}},
\end{eqnarray}

\noindent where $n_{ij} = 2$ for $i \neq j$ and $1$ otherwise.

These $a_{i\pm j}$ represent total flux amplitudes for the combination frequencies and are given in terms of the total flux amplitudes of the parent modes ($a_i$, $a_j$).  Because we measure an integrated flux in a restricted wavelength range, these amplitudes are not analogous to the ones we measure.  However, they can be transformed into quantities like those we measure by integrating over the appropriate spherical harmonic viewed at some inclination ($\Theta_\circ$) in the presence of an Eddington limb-darkening law, and then applying a bolometric correction ($\alpha_\lambda$) appropriate for the detector and filter combination.

Calculating the integrated amplitude requires an expression for the flux in the presence of limb darkening.  For a parent mode, which is assumed to have the angular dependence of a spherical harmonic, Wu gives:

\begin{eqnarray}
\label{glm}
g_\ell^m(\Theta_{\circ})\equiv {{1}\over{2\pi}} \oint_0^{2\pi}d\phi\int_{\pi/2}^0 Re[Y_{\ell}^m(\Theta,\Phi)] \left(1+{{3}\over{2}}\cos(\theta)\right)\cos(\theta)d\cos(\theta),
\end{eqnarray}

\noindent where $(\theta,\phi)$ are in the coordinate system defined by the observer's line of sight, and $(\Theta,\Phi)$ are aligned to the pulsation axis of the star.  These two coordinate systems are separated by the angle $\Theta_\circ$, which is the inclination of the star.  Evaluating this integral requires estimating this inclination, and applying the appropriate coordinate transformation (see Appendix A of Wu).

For the combination frequencies, the integrated flux depends on the product of the spherical harmonics of the parent modes:

\begin{eqnarray}
\label{Glm}
G_{\ell_i~\ell_j}^{m_i \pm m_j}(\Theta_\circ)\equiv {{N_{\ell_i}^{m_i} N_{\ell_j}^{m_j}}\over{2\pi}} \oint_0^{2\pi}d\phi \int_{\pi/2}^0 \rho_{\ell_i}^{m_i}(\Theta)\rho_{\ell_j}^{m_j}(\Theta)\cos{((m_i \pm m_j)\Phi)} \times \nonumber \\ \left(1+{{3}\over{2}}\cos(\theta)\right)\cos(\theta)d\cos(\theta),
\end{eqnarray}

\noindent where the $\rho_\ell^m(\Theta)$ are Legendre polynomials, and the $N_{\ell}^m$ are the normalization factors for the parent mode spherical harmonics.  Our expression differs from that of Wu slightly, in that we explicitly retain these normalization factors.

The bolometric correction is simpler, since it is only a numeric factor expressing the ratio of the amplitudes measured by the detector to the bolometric variations given by the theory.  We calculated this factor using model atmospheres of different temperatures provided by Koester \citep[discussed in][]{fkb97}.  The observations we analyze in this paper are either white light measurement using a bi-alkali photocathode or CCD measurements with a red cutoff filter (BG40).  We applied the known sensitivity curves of these systems and the UV cutoff of the Earth's atmosphere to the model spectra and found bolometric corrections of $\alpha_\lambda = 0.46$ and $0.42$, respectively.  Because of the wavelength dependence of limb darkening these corrections depend upon the value of $\ell$ assumed for the modes, but this dependence is weak at optical wavelengths.  Our values are calculated assuming $\ell=1$.  They are so close to the value that Wu used ($0.4$) that we have decided to retain her value of $0.4$ to make our results directly comparable to hers.

Now we can write the observable flux change at the photosphere in terms of  

\begin{equation}
\label{parentamp}
{\left({{\delta f}\over{f}}\right)_i} = \alpha_\lambda a_i g_{\ell_i}^{m_i}(\Theta_\circ)
\end{equation}

\begin{equation}
\label{comboamp}
\left({{\delta f}\over{f}}\right)_{i\pm j} = \alpha_\lambda a_{i\pm j} G_{\ell_i~\ell_j}^{m_i \pm m_j}(\Theta_\circ)
\end{equation}

\noindent so the predicted combination amplitude is:

\begin{eqnarray}
\label{amplitude}
{\left({{\delta f}\over{f}}\right)_{i\pm j}} =  {{n_{ij}}\over{2}} {{{\left({{\delta f}\over{f}}\right)_i} {\left({{\delta f}\over{f}}\right)_j}}\over{2\alpha_\lambda}} {{\mid 2\beta+\gamma \mid (\omega_i \pm \omega_j)\tau_{c_\circ}}\over{\sqrt{1+[(\omega_i \pm \omega_j)\tau_{c_\circ}]^2}}} {{ G_{\ell_i~\ell_j}^{m_i \pm m_j}(\Theta_\circ)}\over{ g_{\ell_i}^{m_i}(\Theta_\circ) g_{\ell_j}^{m_j}(\Theta_\circ)}}.
\end{eqnarray}

\noindent Equation~\ref{amplitude} is the expression we use to calculate the predicted combination frequency amplitudes for various assumptions of $\ell$ and $m$ for the parent modes. 

We reiterate that the bolometric corrections of the two parent modes are really only equal if they are modes of the same $\ell$.  Moreover, the value for $\alpha_\lambda$ for the combination frequency amplitude in equation~\ref{comboamp} will be a linear combination of the bolometric corrections of the two parent modes.  Consequently, the $1 \diagup \alpha_\lambda$ dependence of $(\delta f \diagup f)_{i\pm j}$ in equation~\ref{amplitude} (and of $R_c$ in equation~\ref{Rc}) is only an approximation.

In addition to $\alpha_\lambda$, calculating a prediction for the combination frequency amplitudes requires six additional quantities, $(\delta f \diagup f)_i$, $\omega_i$, $\beta$, $\gamma$, $\tau_{c_\circ}$, and $\Theta_\circ$.  The first two are the parent mode amplitude and frequency measured from the Fourier transform.  $\beta$ and $\gamma$ are theoretical atmospheric parameters defined by Wu, and $\tau_{c_\circ}$ is the convective timescale estimated from the longest period mode.  The final quantity, $\Theta_\circ$, is the inclination, which we discuss later.  

Physically, it is useful to rearrange equation~\ref{amplitude} into the form:

\begin{eqnarray}
\label{Rc}
R_c\equiv {{\left({{\delta f}\over{f}}\right) _{i\pm j}}\over{n_{ij}{\left({{\delta f}\over{f}}\right)_i} {\left({{\delta f}\over{f}}\right)_j}}} = \left[{{\mid 2\beta+\gamma \mid (\omega_i \pm \omega_j)\tau_{c_\circ}}\over{4\alpha_\lambda\sqrt{1+[(\omega_i \pm \omega_j)\tau_{c_\circ}]^2}}}\right] {{ G_{\ell_i~\ell_j}^{m_i \pm m_j}(\Theta_\circ)}\over{ g_{\ell_i}^{m_i}(\Theta_\circ) g_{\ell_j}^{m_j}(\Theta_\circ)}} \nonumber \\ = F(\omega_i,\omega_j,\tau_{c_\circ},2\beta+\gamma) {{ G_{\ell_i~\ell_j}^{m_i \pm m_j}(\Theta_\circ)}\over{ g_{\ell_i}^{m_i}(\Theta_\circ) g_{\ell_j}^{m_j}(\Theta_\circ)}} \nonumber \\ = {\cal F}~~~{\cal G}.~~~~~~~~~~~~~~~~~~~~~~~~~~~~~~~~~~~~~~~~
\end{eqnarray}

\noindent The ratio $R_c$ is a dimensionless ratio between the combination frequency and the product of its parents, as introduced by \citet{vcw00}.  It is instructive to consider the two terms on the right hand side of equation~\ref{Rc} separately.  The first term (${\cal F}$) incorporates the physics particular to this model, i.e., the thermal properties of the convection zone, while the second term (${\cal G}$) is geometric, and will be present in any theory that accounts for combination frequencies using nonlinear mixing.  In Wu's theory, the $\ell$ and $m$ dependence is entirely contained within this geometric term (except for the $\ell$ dependence of the bolometric correction discussed before).  Thus mode identification is possible if changes in ${\cal G}$ with $\ell$ are large compared to the natural variations in ${\cal F}$.

\begin{figure}
\includegraphics[scale=.7,angle=-90]{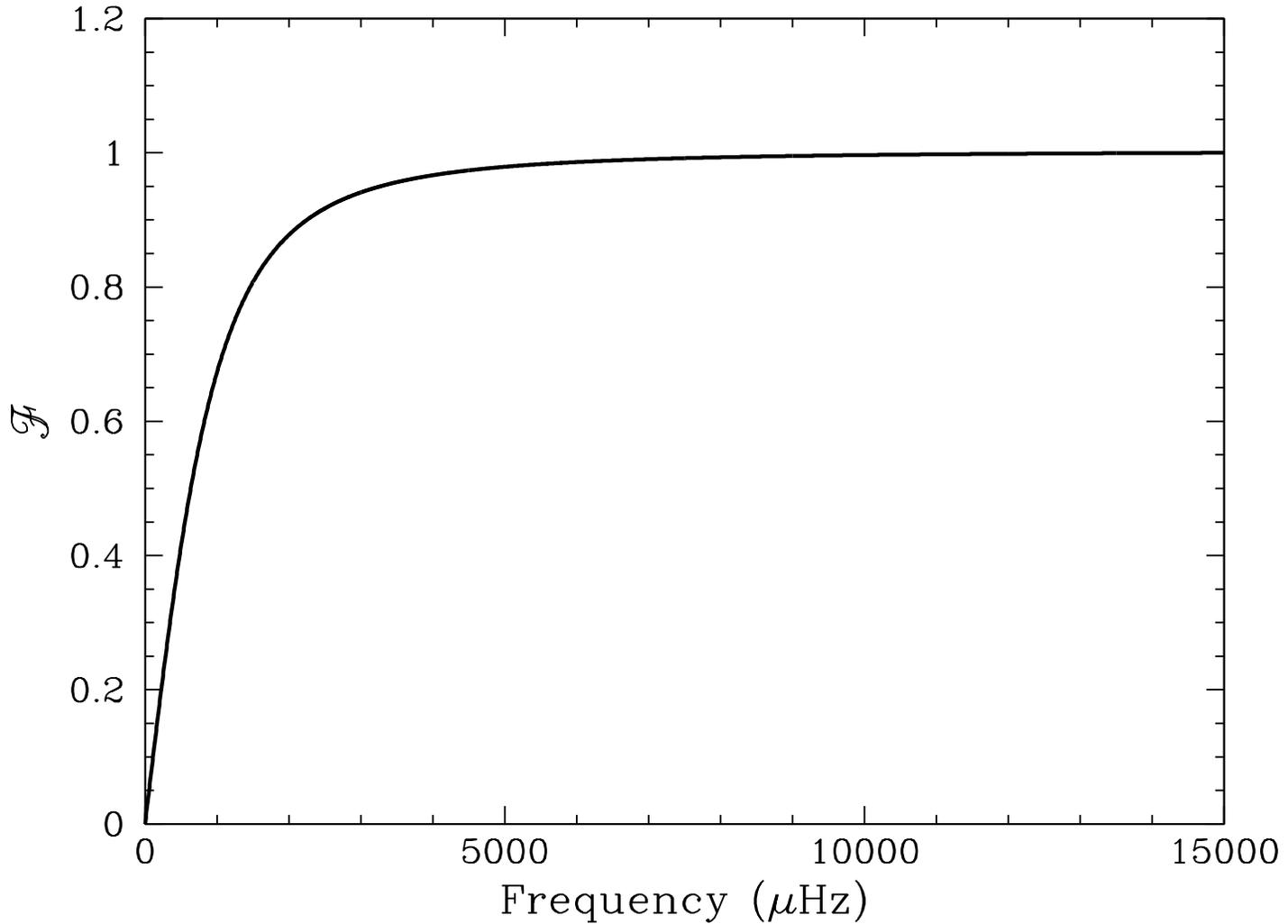}
\caption{Typical ${\cal F}$ dependence on frequency, normalized to one.  ${\cal F}$ incorporates the physics of the model of Wu (see equation~\ref{Rc}).  For a given star, it is only dependent on the frequency of the combination or harmonic.  The only parameter in ${\cal F}$ whose value varies across stars is $\tau_{c_\circ}$, which affects the location of the low frequency roll-off.  The other component of $R_c$, ${\cal G}$, depends upon $\ell$, $m$, and $\Theta_\circ$ (see Figure~\ref{Ggg}). }
\label{F}
\end{figure}

In this respect, the theory of Wu is promising.  For any individual star with multiple pulsation modes, the only parameter in ${\cal F}$ that changes from one mode to another is $\omega$.  Moreover, the functional dependence on $\omega$ is such that for typical ZZ Ceti {\it sum} frequencies the variations in ${\cal F}$ are so small that ${\cal F}$ $\sim$ constant (see Figure~\ref{F}).  The same is not true for difference frequencies, which lie at low frequencies and are therefore suppressed.  For comparison between modes in two different stars, the other parameters in ${\cal F}$ change slowly, so that small adjustments to ${\cal F}$ should be able to reproduce a variety of stars with similar temperature and mean pulsation period, such as the ensemble we consider in this paper.
\clearpage

\clearpage
The theory of BFW can be expressed in the same form as equation~\ref{Rc} by replacing ${\cal F}$ with their tabulated atmospheric model parameters.  However the BFW ${\cal F}$ is independent of pulsation frequency for different modes in any single star, so low frequency difference modes are not suppressed.  For comparisons between modes in different stars, the BFW theory is radically different from Wu's.  The BFW ${\cal F}$ term, which is normally an order of magnitude smaller than in the Wu theory, grows to comparable size for a narrow range of temperature that depends sensitively on stellar mass.  Thus the expectation of the BFW theory is that combination frequencies in {\it most} ZZ Cetis will be smaller than in the theory of Wu (for the same $\ell$).  Moreover, the temperature sensitivity of ${\cal F}$ makes mode identification more problematic if the BFW theory is correct.  Without very precise temperature measurements, it is impossible to distinguish between large differences in ${\cal F}$ arising from temperature differences, and large changes in ${\cal G}$, the geometric term, arising from differences in $\ell$ or $\Theta_\circ$.

\begin{figure}
\epsscale{.8}
\plotone{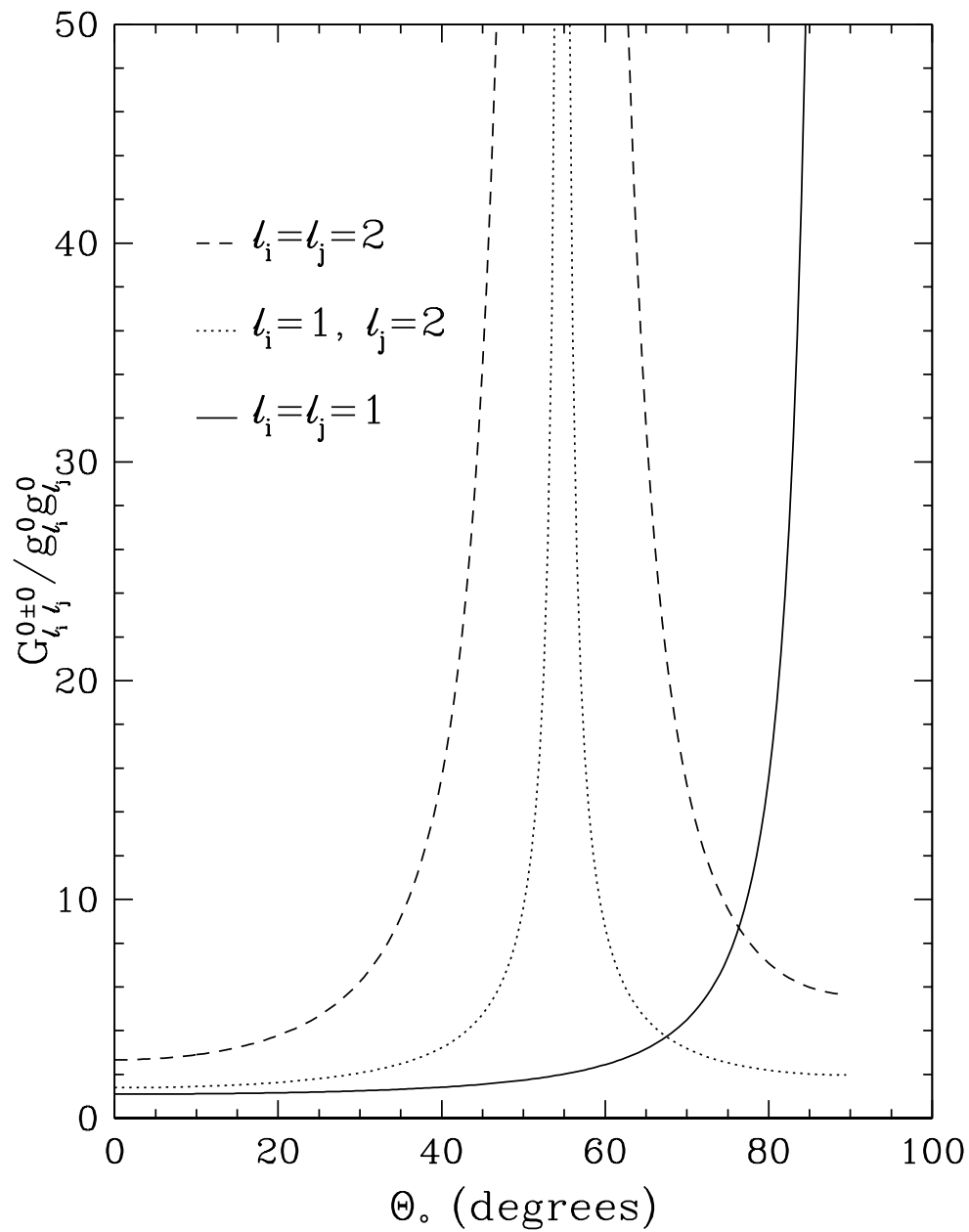}
\caption{${\cal G}$ with $m_i=m_j=0$ (see equation~\ref{Rc}) plotted as a function of inclination angle ($\Theta_\circ$).  For low inclinations ($\Theta_\circ \lesssim 25^\circ$), the predicted amplitudes of the combination frequencies show only a gradual increase with $\ell_i=\ell_j=1$ (solid line), $\ell_i=\ell_j=2$ (dashed line), and $\ell_i=1$, $\ell_j=2$ (dotted line).  }
\label{Ggg}
\epsscale{1.0}
\end{figure}

\begin{figure}
\epsscale{.8}
\plotone{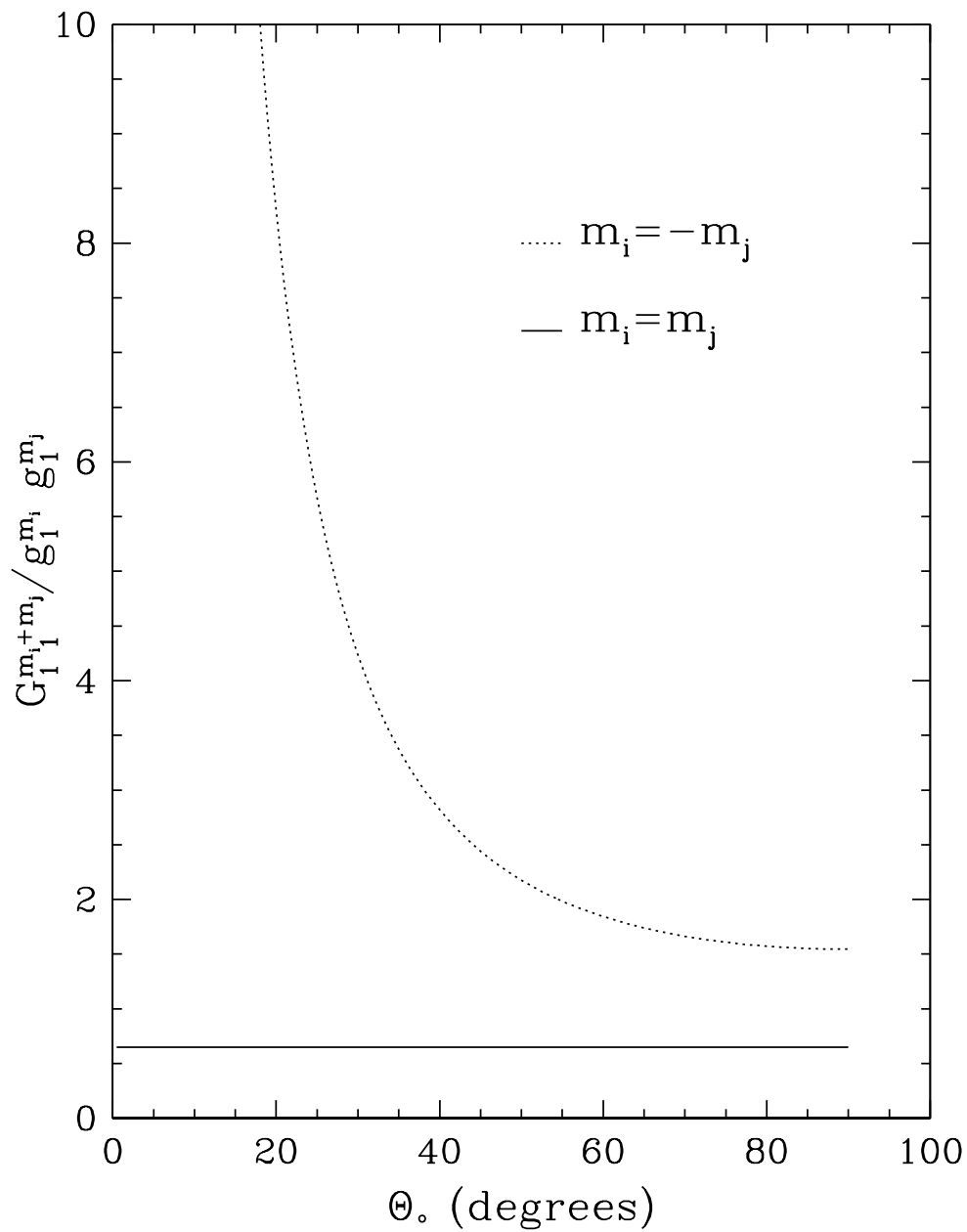}
\caption{${\cal G}$ with $\ell_i=\ell_j=1$ and $m_i+m_j$ (see equation~\ref{Rc}) plotted as a function of inclination angle ($\Theta_\circ$).  The amplitude of the combination frequencies is insensitive to inclination when $\Theta_\circ \gtrsim 60^\circ$ for $m_i=-m_j$ (dotted line).  When both parent modes have the same $m$ (solid line), the combination amplitude is {\it always} independent of inclination.  ${\cal G}$ with $m_i-m_j$ can be obtained by letting the dotted line represent $m_i=m_j$ and the solid line represent $m_i=-m_j$.}
\label{G1111gg}
\epsscale{1.0}
\end{figure}

For either theory, the calculation of ${\cal G}$ in equation~\ref{Rc} requires assigning a value to the inclination of the pulsation axis to our line of sight.  Following \citet{pes85}, we can estimate the inclination for each star by comparing finely split modes of different $m$.  This requires that we make potentially dubious assumptions about the relative intrinsic sizes of pulsation modes, but the final result is not very sensitive to the assumptions.  Figures~\ref{Ggg} and \ref{G1111gg} show why.  Figure~\ref{Ggg} shows the dependence of the geometric factor on inclination for $m=0$ modes.  It varies very slowly over a large range, and then changes rapidly when we look directly down upon a nodal line.  For $\ell=1$, this occurs near $\Theta_\circ = 90^\circ$, because the parent modes are totally geometrically cancelled and the combination frequencies are not.  However, the apparent size of these modes, as opposed to the ratio of their sizes, diminishes rapidly near $90^\circ$, and they eventually fall below the noise threshold of the Fourier transform.  At the same viewing angle, if any $m\neq 0$ modes are present, they will dominate the power spectrum and so will their combination frequencies.  As Figure~\ref{G1111gg} shows, these combinations are not very sensitive to inclination for $\Theta_\circ$ near $90^\circ$.  In other words, the analysis of combination frequencies requires that they be detectable.  At low inclination, only $m=0$ combination frequencies are detectable and at low inclination these are insensitive to $\Theta_\circ$, at high inclinations only $m\neq 0$ modes are detectable and at high inclination these are insensitive to $\Theta_\circ$.  For higher $\ell$ the situation is more complicated, because there are more nodal lines, but the basic argument still applies.

With this in mind, following \citet{pes85}, we have assumed that the intrinsic mode amplitudes are the same for all the modes within a multiplet.  When modes of a specific $\ell$ value are rotationally split into $2\ell+1$ modes, the inclination of a star can be found by equating the amplitude ratio of the $m = 0$ peak and an $m\neq 0$ peak with the corresponding ratio of $N_\ell^m\rho_\ell^m(\Theta_\circ)$ for both values of $m$.  The $N_\ell^m$ are the coefficients of the spherical harmonic, $Y_{\ell}^m(\Theta, \Phi)$, and the $\rho_{\ell}^m(\Theta_\circ)$ are the Legendre Polynomials.  We estimated the inclination for each star by averaging the amplitudes of the $m=\pm 1$ members of the largest amplitude $\ell=1$ multiplets in each star.

There are four stars in our study with detected combination frequencies.  For three of these, the \citet{pes85} method yields low inclination ($\Theta_\circ < 20^\circ$).  The only combination frequencies detected in these stars are combinations of $m=0$ parent modes, as established by their singlet nature or by their central location in a frequency symmetric triplet.  Figure~\ref{Ggg} shows that except for at large inclination, the amplitudes of the combination frequencies are not very sensitive to inclination for the central, $m=0$, parent modes.  The fourth star (GD 244) has a high inclination ($\Theta_\circ \gtrsim 80^\circ$), and shows only combinations of $m\neq 0$ parent modes.  Figure~\ref{G1111gg} shows that at high inclinations the amplitudes of combination frequencies are not very sensitive to inclination when the parent modes are $m=\pm 1$ members of an $\ell=1$ triplet.  In fact, for certain $m$ combinations, the amplitudes of combination frequencies are measured {\it independent} of inclination.  Therefore, for all stars that we analyze, the amplitudes of the combination frequencies are at most weakly dependent on the inclination, as long as the value of $\ell$ is small.  Hence, the approximation of inclination is a small source of error in our analysis.

With independent estimates of inclination, the only factor that remains unknown in the factor ${\cal G}$ of the theory of Wu is the value of $\ell$ for each mode.  Thus we can compare the measured combination frequencies in the data, if any, to the predicted amplitudes of combination frequencies under various assumptions for the $\ell$ value of the parent modes.  In this way we can hope to constrain or actually measure the value of $\ell$.  We will see that harmonics of a single mode are more valuable in this enterprise than combinations between two different modes.  This is because there is a greater contrast in the theory between same-$\ell$ combinations, and for harmonics there is only one parent, and therefore only one $\ell$ involved.  In the section that follows, we apply the theory to eight low amplitude hot ZZ Ceti stars, and show that it is possible to establish the values of $\ell$ for most modes in these stars.  
\clearpage

\clearpage
\section{DATA REDUCTION AND ANALYSIS}
\label{data}

The data we present in this section are a combination of published Fourier spectra, new reductions of archival Whole Earth Telescope data, and original data obtained with the McDonald Observatory 2.1-m Struve telescope.  They constitute all of the currently-available data with frequency resolution sufficient to measure amplitudes or amplitude limits for combination frequencies in hot DAV stars.  The purpose of this section is to extract mode frequencies and amplitudes via Fourier methods.  In principle, phase measurements are also possible, but in practice they are too noisy to be useful.  In cases where combination frequencies are not sufficiently above the noise level to detect, we record the upper limit for comparison to theory.

\subsection{Data Reduction}

\subsubsection{Published Data}
\label{published}

The stars for which we use published data include G117-B15A, G185-32, GD 165, R548, and G226-29 \citep{kep95b,cas04,ber93,muk03,kep95a}.  These stars were all included as secondary target stars in Whole Earth Telescope campaigns \citep[WET;][]{nat90} and the referenced papers present these WET data.  \citet{ber93} (GD 165) and \citet{muk03} (R548) also include Canada-France-Hawaii Telescope (CFHT) observations, and \citet{muk03} includes supplemental McDonald Observatory observations.  All of the published data we have analyzed for combination frequencies are found in these publications, along with explanations of the reduction and analysis procedures.

\subsubsection{Unpublished Data and New Reductions}

\label{unpublished}

We obtained time-series photometry data on both GD 66 and GD 244 in 2003 and 2004 with the McDonald Observatory 2.1-m Struve telescope using the prime-focus ARGOS CCD photometer with a BG40 Schott glass filter \citep{nm04}.  We observed GD 66 on fourteen nights during three observing runs totaling 155,130 seconds of data as indicated in the Journal of Observations, Table~\ref{j_obsGD}.  We used two integrations times (10 s for the 2003 October run and 5 s for the 2003 November and 2004 January runs, see Table~\ref{j_obsGD}).  To combine the runs into one lightcurve we binned the 5 s observations into 10 s bins.  We observed GD 244 on ten nights during three observing runs totaling 124,500 seconds of data (see Table~\ref{j_obsGD}).  We used an integration time of 5 s for all of our GD 244 observations.  We performed a complete reduction of the original data for GD 66 and GD 244 using the methods described by \citet{muk04}. 
\clearpage
\begin{deluxetable}{llllc}
\tablecolumns{5}
\tabletypesize{\scriptsize}
\tablewidth{0pt}
\tablecaption{Journal of Observations for GD 66 and GD 244\label{j_obsGD}}
\tablehead{
\colhead{Run Name} & \colhead{Date} & \colhead{Start Time} & \colhead{Length} & \colhead{Integration Time} \\
\colhead{} & \colhead{(UT)} & \colhead{(UT)} & \colhead{(sec)} & \colhead{(sec)} \\
\cline{1-5} \\
\multicolumn{5}{c}{GD 66}} 
\startdata
A0726 &2003 Oct 25 &9:04:23 &12750 &10 \\

A0729 &2003 Oct 27 &9:42:27 &1870 &10 \\
A0730 &2003 Oct 27 &10:58:12 &5420 &10 \\
A0733 &2003 Oct 28 &6:52:39 &19530 &10 \\
A0738 &2003 Oct 29 &11:08:30 &4560 &10 \\
A0742 &2003 Oct 31 &9:32:33 &10720 &10 \\
A0746 &2003 Nov 1 &8:17:12 &3970 &10 \\
A0755 &2003 Nov 19 &7:29:14 &11475 &5 \\
A0767 &2003 Nov 22 &5:11:34 &14785 &5 \\
A0789 &2003 Nov 29 &6:08:55 &12045 &5 \\
A0793 &2003 Nov 30 &6:49:05 &21585 &5 \\
A0795 &2003 Dec 1 &5:44:42 &11940 &5 \\
A0835 &2004 Jan 20 &3:26:37 &11715 &5 \\
A0838 &2004 Jan 21 &3:09:07 &12765 &5 \\
\cutinhead{GD 244}
A0693 &2003 Sep 2 &8:42:35 &10270 &5 \\
A0695 &2003 Sep 3 &4:21:55 &7705 &5 \\
A0700 &2003 Sep 4 &5:23:28 &9555 &5 \\
A0705 &2003 Sep 5 &4:50:31 &15070 &5 \\
A0732 &2003 Oct 28 &1:30:10 &18770 &5 \\
A0734 &2003 Oct 29 &0:53:42 &12900 &5 \\
A0743 &2003 Nov 1 &0:58:45 &10455 &5 \\
A0766 &2003 Nov 22 &0:54:49 &14880 &5 \\
A0772 &2003 Nov 24 &0:58:49 &12995 &5 \\
A0775 &2003 Nov 25 &1:10:11 &11900 &5 \\
\enddata
\tablecomments{All observations were made with the McDonald Observatory 2.1-m Struve telescope.}
\end{deluxetable}
\clearpage
The data for L19-2 were obtained as the secondary target for the WET campaign XCov 12 in 1995 April \citep[see][]{sul95}.  The observations of L19-2 that were included in this reduction are listed in the Journal of Observations, Table~\ref{j_obsL192}.  The integration times of most runs were 10 s.  We binned the 5 s Mt. John Observatory (MJUO) observations into 10 s bins.  We performed a complete reduction of the original data using the methods described by \citet{nat90} and \citet{kep93}.
\clearpage
\begin{deluxetable}{lllllc}
\tabletypesize{\scriptsize}
\tablewidth{0pt}
\tablecaption{L19-2 Journal of Observations \label{j_obsL192}}
\tablehead{
\colhead{Run Name} & \colhead{Telescope} & \colhead{Date} & \colhead{Start Time} & \colhead{Length} & \colhead{Integration Time} \\
\colhead{} & \colhead{} & \colhead{(UT)} & \colhead{(UT)} & \colhead{(sec)} & \colhead{(sec)}} 
\startdata
S5843 &SAAO 0.75 m &1995 Apr 25 &17:53:00 &5590 &10 \\

S5844 &SAAO 0.75 m &1995 Apr 25 &22:33:00 &5040 &10 \\
RO064 &Itajuba 1.60 m &1995 Apr 25 &23:23:20 &29080 &10 \\
AP2695-1 &MJUO 1.0 m &1995 Apr 26 &11:04:00 &9245 &5 \\
AP2695-2 &MJUO 1.0 m &1995 Apr 26 &13:51:50 &16135 &5 \\
RO065 &Itajuba 1.60 m &1995 Apr 27 &3:53:20 &11810 &10 \\
S5845 &SAAO 0.75 m &1995 Apr 27 &17:23:00 &31900 &10 \\
AP2895 &MJUO 1.0 m &1995 Apr 28 &11:38:20 &18550 &5 \\
S5846 &SAAO 0.75 m &1995 Apr 28 &17:25:00 &24360 &10 \\
RO066 &Itajuba 1.60 m &1995 Apr 28 &22:25:00 &9410 &10 \\

RO067 &Itajuba 1.60 m &1995 Apr 29 &1:59:00 &12600 &10 \\
AP2995 &MJUO 1.0 m &1995 Apr 29 &7:27:10 &39270 &5 \\
S5847 &SAAO 0.75 m &1995 Apr 29 &17:20:00 &31580 &10 \\
RO068 &Itajuba 1.60 m &1995 Apr 29 &22:26:40 &33710 &10 \\
S5848 &SAAO 0.75 m &1995 Apr 30 &21:00:00 &14340 &10 \\
RO069 &Itajuba 1.60 m &1995 Apr 30 &22:11:40 &34560 &10 \\
MY0195 &MJUO 1.0 m &1995 May 1 &7:04:30 &40660 &5 \\
S5849 &SAAO 0.75 m &1995 May 1 &23:41:00 &7890 &10 \\
RO070 &Itajuba 1.60 m &1995 May 2 &1:27:50 &9630 &10 \\
RO071 &Itajuba 1.60 m &1995 May 2 &5:02:20 &10340 &10 \\
DB001 &SAAO 0.75 m &1995 May 2 &18:19:30 &26990 &10 \\
DB002 &SAAO 0.75 m &1995 May 3 &2:38:40 &5090 &10 \\
DB003 &SAAO 0.75 m &1995 May 3 &17:17:20 &29990 &10 \\
RO073 &Itajuba 1.60 m &1995 May 4 &1:20:50 &21510 &10 \\
DB004 &SAAO 0.75 m &1995 May 4 &1:47:20 &8290 &10 \\
DB005 &SAAO 0.75 m &1995 May 4 &19:58:30 &21150 &10 \\
DB006 &SAAO 0.75 m &1995 May 5 &2:02:40 &3200 &10 \\
\enddata
\tablecomments{All data come from the WET campaign XCov12.}
\end{deluxetable}
\clearpage
\subsection{Analysis}
\label{overview}

Of the stars that we studied, four exhibit detectable combination frequencies: GD 66, GD 244, G117-B15A, and G185-32.  The remainder, L19-2, GD 165, R548, and G226-29, do not show combination frequencies, though we will show that Wu's theory suggests that they must be just below the current noise limits.  The temperature and $\log g$ for each star from \citet{ber04} are listed in Table~\ref{tbl-starsTg}.  In \S\ref{detections} and \S\ref{nodetection}, we will present the individual analyses.  We will describe our calculations of the inclination of each star and compare the observed combination frequency amplitudes to the predictions of Wu's theory.
\clearpage
\begin{deluxetable}{lccc}
\tabletypesize{\scriptsize}
\tablewidth{0pt}
\tablecaption{Stellar Information\label{tbl-starsTg}}
\tablehead{
\colhead{Star} & \colhead{$T_{eff}$} & \colhead{$\log g$} & \colhead{Reference} \\
\colhead{} & \colhead{(K)} & \colhead{(cgs)} & \colhead{}} 
\startdata
GD 66 &11,980 &8.05 &1\\
G117-B15A &11,630 &7.97 &1\\
GD 244 &11,680 &8.08 &1\\
G185-32 &12,130 &8.05 &1\\
L19-2 &12,100 &8.21 &1\\
GD 165 &11,980 &8.06 &1\\
R548 &11,990 &7.97 &1\\
G226-29 &12,460 &8.28 &1\\
\enddata
\tablerefs{(1) \citet{ber04}.}
\end{deluxetable}
\clearpage
\subsubsection{Stars With Detected Combination Frequencies}
\label{detections}

\subsubsubsection{\it{GD 66}}
\label{gd66section}

Apart from an analysis by \citet{fon85}, little progress toward understanding GD 66 has been made since \citet{dvc83} first reported its discovery.  The relatively high number of combination frequencies identified in the Fourier transform of GD 66 make it an ideal star to include in this paper. 

\begin{figure} 
\includegraphics[scale=.7,angle=-90]{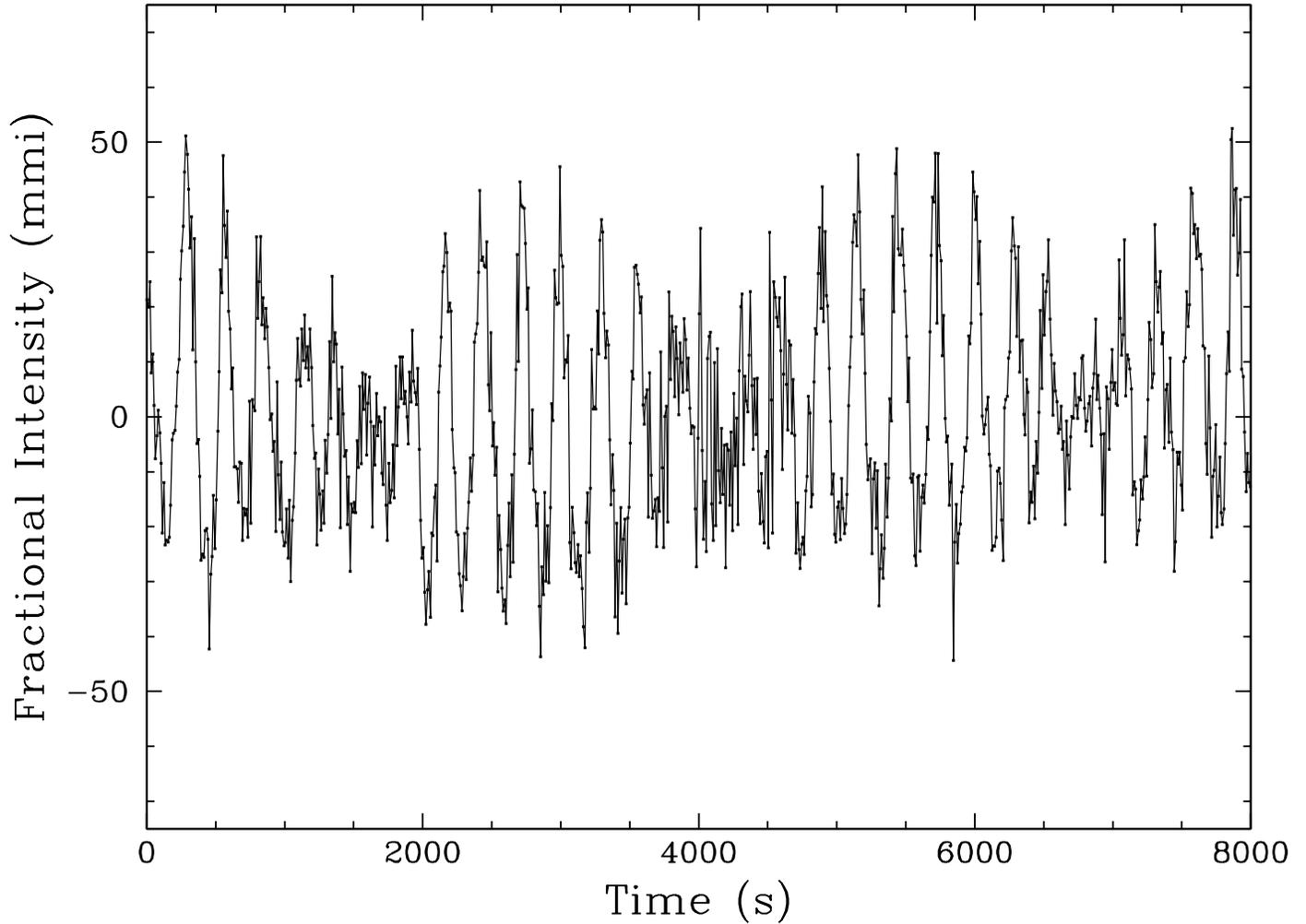}
\caption{Lightcurve of GD 66.  These data were acquired with the ARGOS CCD photometer on the McDonald Observatory 2.1-m Struve telescope with an exposure time of 10 s. }
\label{GD66lc}
\end{figure}

To identify the pulsation modes and combination frequencies of GD 66, we computed a Fourier transform from the reduced and combined lightcurves listed in Table~\ref{j_obsGD}.  We have included a sample lightcurve in Figure~\ref{GD66lc} and the Fourier transform of all GD 66 data in Figure~\ref{GD66ft}.  To identify closely spaced modes in the regions of obvious excess power, we utilized a prewhitening technique similar to that of \citet{ow82} using an iterative nonlinear least squares procedure.  For each peak, we fitted the frequency, amplitude, and phase and then subtracted the fit from the original lightcurve.  We then fitted a second frequency to the altered data, choosing in every case the largest remaining peak, and used the result of this fit to conduct a simultaneous least squares fit to the original data.  Thus at each step in the prewhitening, the frequencies removed are from a simultaneous fit to the original data. 

The problems with applying such a procedure to a highly aliased data set are well known \citep{nat90}, and we have no illusions that we can successfully measure the correct frequencies of the smaller modes in the presence of the contaminating window function.  Nonetheless, the exercise provides two pieces of information that are valuable and reliable.  It tells us how many modes are required to model the data, and gives us crude amplitudes for the members of the multiplet that are useful in estimating inclination.
 \clearpage

\begin{figure} 
\caption{Fourier transform of GD 66.  This FT includes all individual nights of the data that are listed in Table~\ref{j_obsGD}.  We indicate the five pulsation modes and nine identified combination frequencies that we reference in Table~\ref{tbl-GD66}.  We believe the peak that we call F6 is due to the guide error of the telescope.  }
\label{GD66ft}
\end{figure}
\clearpage
In Figure~\ref{GD66_271ft} we show the deconstruction of the 271 second pulsation mode (F1) by prewhitening.  The Fourier transform (FT) of the reduced data in the region of F1 is shown in Figure~\ref{GD66_271ft}a.  The window function, in Figure~\ref{GD66_271ft}b, is the FT of a lightcurve of a sinusoid with the same period, amplitude, and phase as the highest amplitude peak in the original FT that has been sampled in the same manner as the original data.  Figure~\ref{GD66_271ft}c shows the FT of the lightcurve with the largest peak removed.  Figures~\ref{GD66_271ft}c and \ref{GD66_271ft}d are prewhitened FTs that reveal additional low amplitude signals which were previously hidden in the window function of the highest amplitude peak.  Figure~\ref{GD66_271ft}e, in which there is no remaining signal, is an FT of a lightcurve with the three highest amplitude peaks fitted and removed from the original lightcurve.  It is gratifying that the frequencies identified by this deconstruction form a frequency-symmetric triplet, but better sampling will be required to measure all three frequencies with confidence. 

A similar procedure showed that F2 was consistent with a single frequency, to within the noise, and that F3 is a combination of at least three frequencies.  F4 and F5 both show residuals after prewhitening by one frequency, but they are too close to the noise level to deconstruct further.  F6 is at the frequency expected from the drive of the telescope, and a peak at this location is present in all of the ARGOS data, so we do not interpret it as being astrophysically significant.
\clearpage
\begin{figure} 
\caption{ Deconstruction of F1 in GD 66.  a: A Fourier transform of GD 66 near 271 s (F1).  b: A window function obtained by taking an FT of a single sinusoid (with the same period and amplitude of the 271.71 s peak) that has been sampled in the same manner as the data.  c: An FT near F1 with a period of 271.71 s removed.  d: An FT near F1 with periods of 271.71 and 272.20 s removed.  e: An FT near F1 with periods of 271.71, 272.20, and 271.23 s removed. }
\label{GD66_271ft}
\end{figure}
\clearpage
The five dominant pulsation modes (F1~-~F5) and associated power derived from prewhitening are listed in Table~\ref{tbl-GD66} and presented visually in Figure~\ref{gd66periodogram}.  The amplitude of the smallest of our five modes is about three times above $\sqrt{\langle P \rangle}$, where $\langle P \rangle$ is the average power of the FT, yielding a false alarm probability \citep{hb86} of about 20\%.  That is, there is a 20\% chance that F5 is an artifact of noise.  However, the existence of a combination frequency at F1+F5 adds confidence to this detection.  All the pulsation modes except F5 were also identified by \citet{fon01}.  There is a large peak near the location of F5 in their data, but it was not formally significant against the noise level of their FT.

In addition to the five modes detected, we identified nine combination frequencies that were consistently present throughout the three month span of our GD 66 observations, with one exception (F1+F4 was not identified in the 2003 October data set).  We verified our identification of combination frequencies with a computer program inspired by \citet{klein95}.  Our program allows a user to select pulsation modes in the data, calculates all possible combination frequencies of these modes, and then searches for significant combination frequencies in the data.  The search is conducted over an estimated error range equal to the resolution of the Fourier transform.  We consider this a better estimate of the frequency error than the smaller values from the least squares fit because the combination frequencies are usually low amplitude, i.e., only a few times as large as the background.  The frequencies of such small signals are pulled by the presence of unresolved noise peaks, while the least squares fit formally assumes only a single unblended frequency is present. 
  The program successfully identified all combination frequencies that we found by inspection in the data, and also found some combinations that we had missed in our visual search.  In some cases, the highest amplitude peak among the combination frequency and its aliases did not fall within the error range we established.  In these cases, prewhitening a forced fit of the expected combination frequency successfully removed the signal, and we have listed the amplitude from the forced frequency fit.  Our listed period errors do not include the uncertainty in identifying the true peak among the alias peaks and these dominate the error for combination frequencies.
\clearpage
\begin{deluxetable}{rlllccrrr}
\tabletypesize{\scriptsize}
\tablewidth{0pt}
\tablecaption{GD 66 Periods and Mode Identifications \label{tbl-GD66}}
\tablehead{

\colhead{Mode Label} & \colhead{Frequency} & \colhead{Period} &\colhead{$\sigma_p$} & \colhead{Amplitude} &\colhead{$\sigma_{amp}$} & \colhead{$\Delta f$ \tablenotemark{a}} & \colhead{$\ell$} & \colhead{$m$ \tablenotemark{b}} \\
\colhead{} & \colhead{($\mu Hz$)} & \colhead{(sec)} & \colhead{(sec)} & \colhead{(mma)} & \colhead{(mma)} & \colhead{($\mu Hz$)} & \colhead{} & \colhead{}} 
\startdata
F1 &3673.753 &272.2012 &0.0004 &2.50 &0.17 &-6.582? &1 &-1 \\
\nodata &3680.335 &271.7144 &0.0001 &16.70 &0.16 &\nodata &1 &0 \\
\nodata &3686.927 &271.2286 &0.0004 &2.93 &0.17 &6.592? &1 &+1 \\
F2 &3302.889 &302.7653 &0.0001 &11.29 &0.19 &\nodata &1 &0 \\
F3 &5049.227 &198.0501 &0.0002 &2.65 &0.21 &-10.355? &1 &-1 \\
\nodata &5059.582 &197.6448 &0.0001 &4.21 &0.21 &\nodata &1 &0 \\
\nodata &5070.3 \tablenotemark{c} &197.23 &\nodata &1.77 &\nodata &10.7? &1 &+1? \\
F4 &3902.680 &256.2341 &0.0004 &2.48 &0.21 &-5.603? &1? &-1? \\
\nodata &3908.283 &255.8668 &0.0003 &3.43 &0.21 &\nodata &1? &0? \\	
F5 &1911.121 &523.2533 &0.0016 &2.33 &0.22 &\nodata &1 or 2 &? \\
\nodata &1928.257 &518.6029 &0.0021 &1.77 &0.22 &17.136? &1 or 2 &? \\
F6 \tablenotemark{d} &8127.845 &123.0338 &0.0002 &1.30 &0.22 &\nodata &guide &\nodata \\
2F1 &7360.670 &135.8572 &0.0002 &1.57 &0.22 &0.001 &\nodata &\nodata \\
F1+F2 &6983.219 &143.2004 &0.0001 &2.83 &0.21 &0.005 &\nodata &\nodata \\
F1+F3 &8747.254 &114.3216 &0.0002 &0.88 &0.22 &-7.336 &\nodata &\nodata \\
F1+F4 &7588.604 &131.7765 &0.0004 &0.64 &0.22 &0.015 &\nodata &\nodata \\ 
F1+F5 &5587.599 &178.9678 &0.0007 &0.61 &0.22 &3.857 &\nodata &\nodata \\
2F2 &6605.802 &151.3821 &0.0004 &0.80 &0.22 &-0.025 &\nodata &\nodata \\
F2+F3 &8362.479 &119.5818 &0.0004 &0.53 &0.22 &-0.009 &\nodata &\nodata \\
F2+F4 &7210.499 &138.6867 &0.0006 &0.48 &0.22 &0.672 &\nodata &\nodata \\
2F1+F2 &10675.6 &93.6720 &0.0002 &0.52 &0.22 &-11.994 &\nodata &\nodata \\
\enddata
\tablenotetext{a}{For pulsation modes, $\Delta f$ is the separation between the modes in the multiplets and the $m=0$ member.  For combination frequencies, $\Delta f$ is the frequency difference between the calculated and observed combination frequency (i.e., $\Delta f =F1+F2-[F1+F2]$).}
\tablenotetext{b}{The $m$ identifications in the table are based on the frequency splitting alone, not on the size of the combination peaks. }
\tablenotetext{c}{We were unable to obtain a simultaneous fit with this frequency and the other two frequencies in the F3 triplet, though it does seem to be a significantly high amplitude peak above the noise.}
\tablenotetext{d}{F6 is a formally significant peak that is due to the guide error of the telescope and probably does not represent a pulsation mode originating at the star.}
\end{deluxetable}
\clearpage
\begin{figure}
\includegraphics[scale=.7,angle=-90]{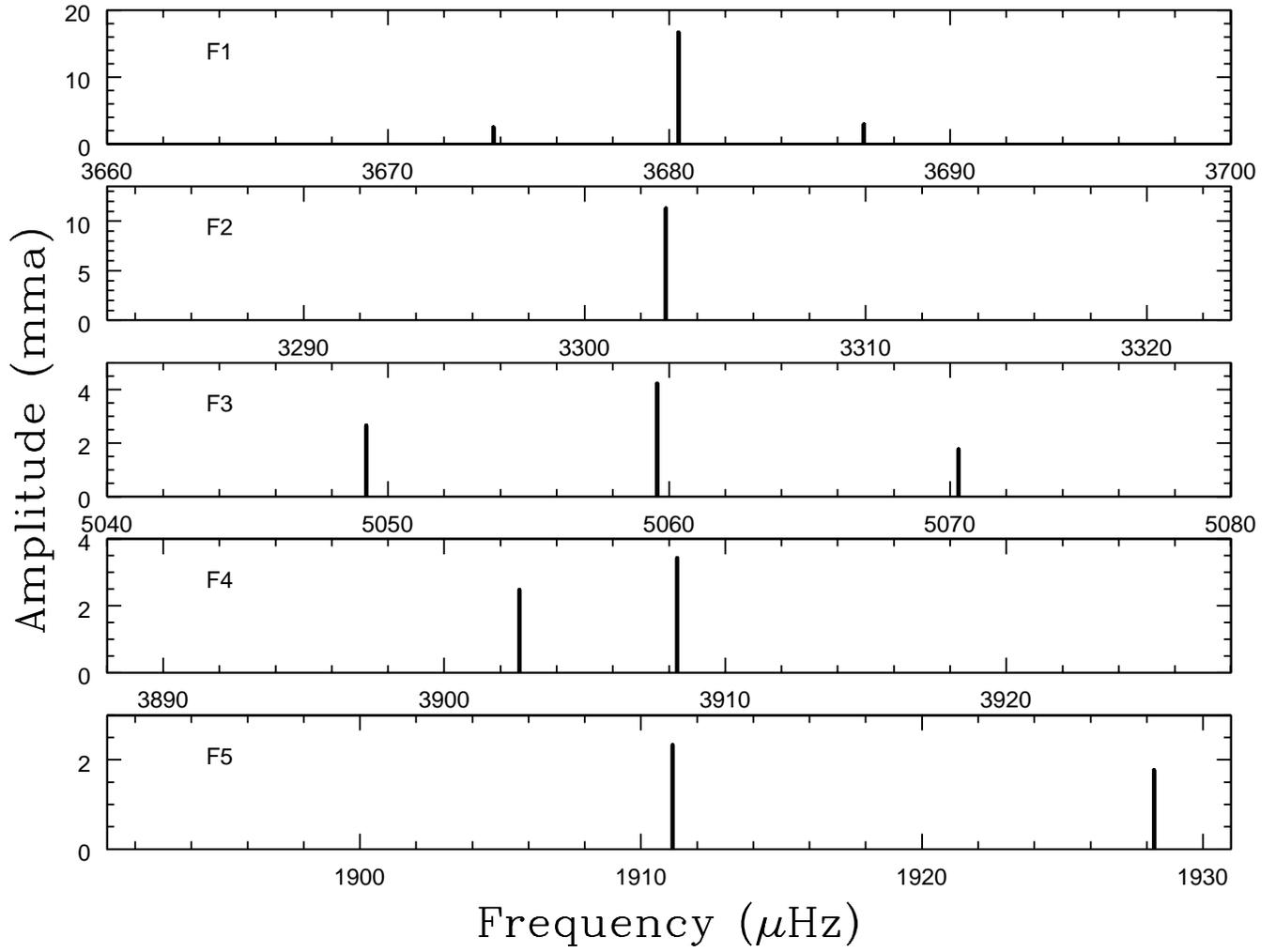}
\caption{ Prewhitened peaks of parent modes in the GD 66 Fourier transform. }
\label{gd66periodogram}
\end{figure}
\clearpage
Having identified and measured the amplitudes of the combination frequencies allows us to calculate the ratio of combination to parent mode amplitudes ($R_c$, see equation~\ref{Rc}) and compare it to the theoretical predictions.  Figure~\ref{RcGD66} shows the observed $R_c$ for all of the detected  two-mode combination frequencies in GD 66.  Where no combination was detected, we have plotted a limit equal to the 1 sigma noise level in the Fourier transform.  The observed $R_c$ do not depend on any theory.

The theoretical calculations of $R_c$ require that we supply an inclination estimate, an estimate of $\tau_{c_\circ}$, and a value for the parameter $2\beta+\gamma$.  For the first, a fit to the largest multiplet using the technique of \citet{pes85} described in \S\ref{theory} yielded $\Theta_\circ=13$ degrees, and we have used that value for all of the theoretical calculations.  The result is not very sensitive to this parameter as long as $\Theta_\circ \lesssim 20^\circ$ (see Figure~\ref{Ggg}).  The inclination calculation requires an assumption for the $\ell$ identification of F1.  However, this calculation is scarcely sensitive to our $\ell=1$ assumption for F1.  If F1 is actually an $\ell=2$ mode, then the inclination of the star is 8 degrees, and still in the range where our results are insensitive to inclination.  For $\tau_{c_\circ}$, we have used the value of 523 s, which is the longest period in GD 66, for reasons discussed in \S\ref{theory}.  This value only affects the location of the frequency roll-off, not the predicted combination frequency amplitudes in the high frequency limit.  Finally, because the detected combinations have similar values of $R_c$, indicating similar $\ell$, we decided to treat $2\beta+\gamma$ in this star as a free parameter.  The solid line in Figure~\ref{RcGD66} shows the best fit under the assumption that all modes are $\ell=1$.  The fitted value of $2\beta+\gamma$ is $-9.35$, very close to the value Wu herself used ($-10$) for her comparison to G29-38, and to her theoretically calculated value ($-12.6$).  Any other assumption for $\ell$ would yield values of $2\beta+\gamma$ different by a factor of $\sim 3$ or more, and we do not consider this a reasonable possibility.  Using $\ell=1$, the detected combination frequencies fit the model with a reduced $\chi^2$ of 0.78, which may indicate that we have slightly overestimated our errors.  We used the errors from the least square fits, which are often regarded as underestimates \citep{win91}, but we see no evidence for that here.
\clearpage
\begin{figure}
\includegraphics[scale=.7,angle=-90]{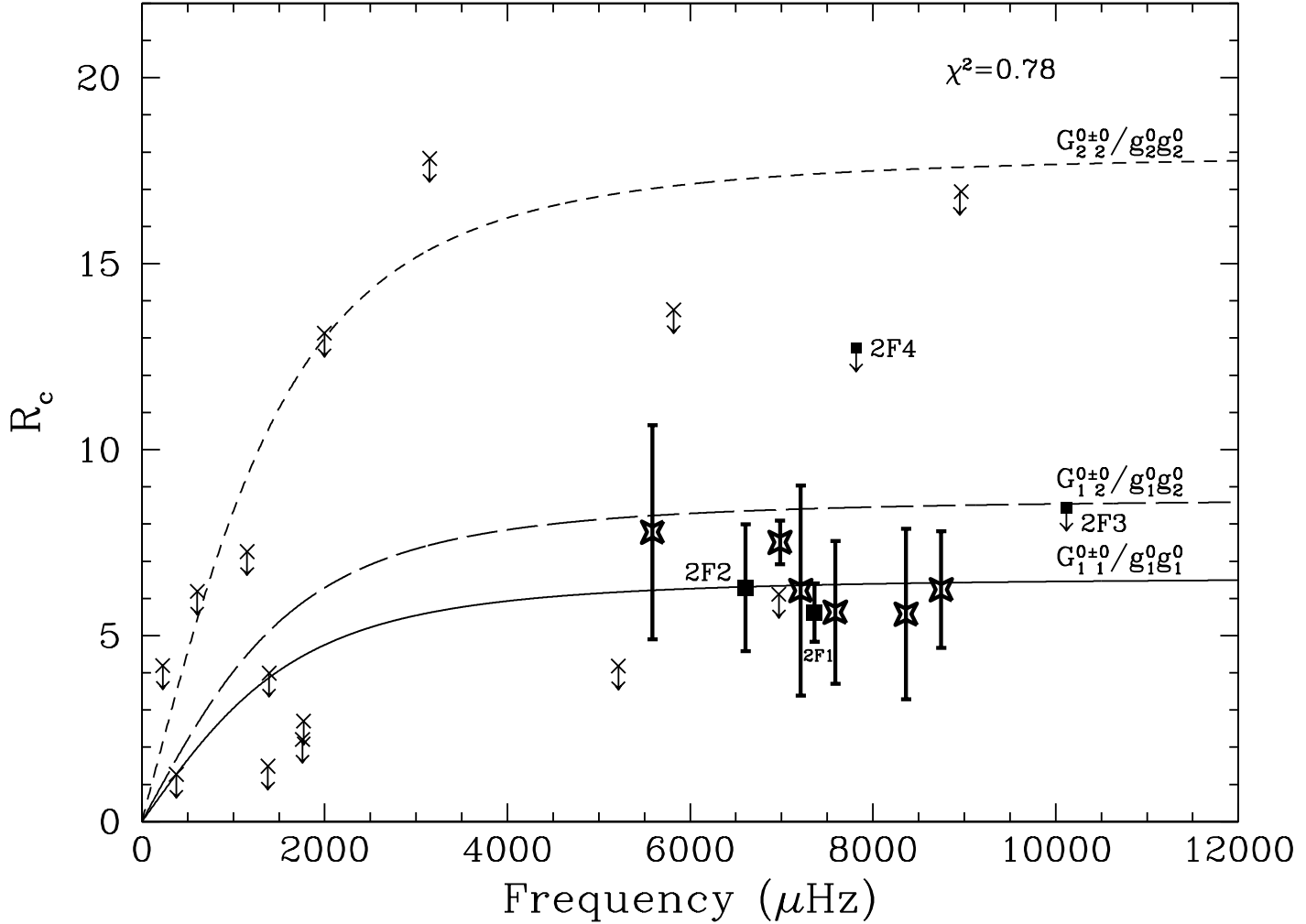}
\caption{Ratio of combination to parent mode amplitudes ($R_c$) for GD 66.  The lines are theoretical predictions for $G_{1~1}^{0+0} \diagup g_1^0 g_1^0 (\Theta_\circ=13^\circ)$ (solid line), $G_{1~2}^{0+0} \diagup g_1^0 g_2^0 (\Theta_\circ=13^\circ)$ (long-dashed line), and $G_{2~2}^{0+0} \diagup g_2^0 g_2^0 (\Theta_\circ=13^\circ)$ (dashed line).  The data points are the detected harmonics or limits (filled squares), detected cross combination frequencies (stars), and limits for the cross combinations (crosses).  The downward arrows on the limits indicate that the points represent maximum values.  The upper limit for $R_c$ of 2F5 lies at 124 (not shown). }
\label{RcGD66}
\end{figure}
\clearpage
Even though the data show tight scatter about the line representing combinations between two $\ell=1$ modes, we cannot conclude that all of the modes are $\ell=1$ on that basis alone.  The line representing $\ell=1,2$ combinations (long dashes in Figure~\ref{RcGD66}) is  close to the $\ell=1,1$ line, and falls within the error bars for some combinations.  Fortunately, the theoretical lines for same-$\ell$ combinations are well-separated, suggesting that harmonics, which are same-$\ell$ by definition, might be able to constrain $\ell$ when cross combinations (i.e., all combinations that are not harmonics) cannot.  We have detected harmonics for F1 and F2, and measured limits for the harmonics of the remaining three modes.  These are shown in Figure~\ref{RcGD66} as filled squares.  The measured $R_c$ for F1 and F2 and the limits for F3 and F4 are only consistent with $\ell=1$, and so we identify all of those modes as $\ell=1$, F4 somewhat tentatively because the limit is not very stringent.  The limit for F5 is too large ($R_c=124$) to fit on our plot and does not constrain $\ell$ uniquely.  Though our identification relies primarily on harmonics, the cross combinations are all consistent with this conclusion.  The $m$ identifications we assigned in Table~\ref{tbl-GD66} arise from frequency splitting only, and are not derived from combination frequency amplitudes.

Finally, the $R_c$ limits we have plotted at low frequency suggest that the roll-off expected from the theory actually occurs, which eliminates from consideration competing models, including the BFW theory, that predict no frequency dependence for $R_c$.

\subsubsubsection{\it{GD 244}}

\citet{fon01} first reported the detection of variability in the DAV white dwarf GD 244, but no subsequent observations have been published.  The Fourier transforms of GD 244 and GD 66 both contain large pulsation modes near 200, 256, and 300 seconds.  The similarity of GD 244 to GD 66, including its relatively high number of combination frequencies, makes it another ideal candidate to include in this study.

\begin{figure}
\includegraphics[scale=.7,angle=-90]{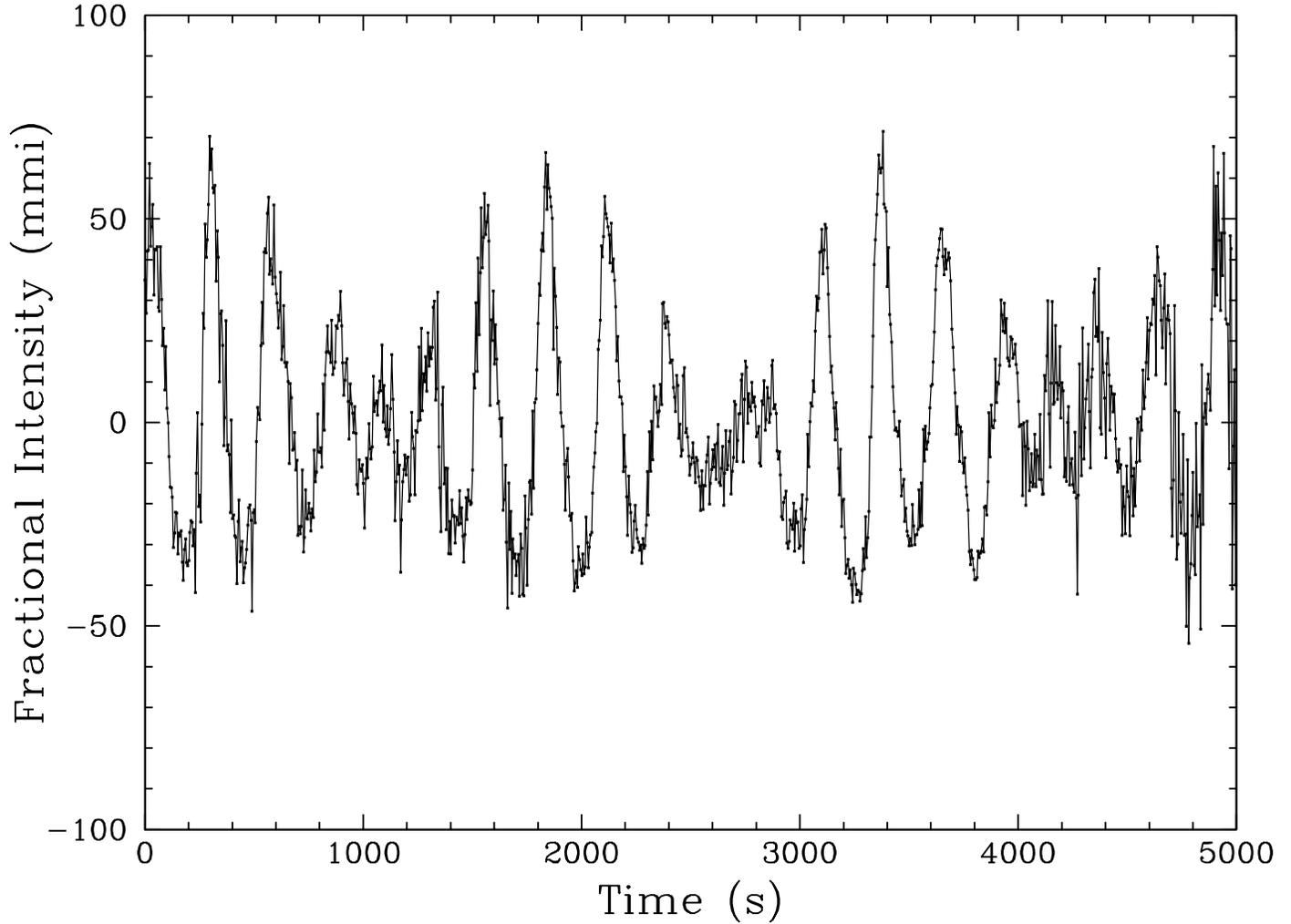}
\caption{Lightcurve of GD 244.  These data were acquired with the ARGOS CCD photometer on the McDonald Observatory 2.1-m Struve telescope with an exposure time of 5 s. }
\label{GD244lc}
\end{figure}

To identify the pulsation modes and combination frequencies of GD 244, we computed a Fourier transform from the reduced and combined lightcurves listed in Table~\ref{j_obsGD}.  We have included a sample lightcurve in Figure~\ref{GD244lc} and the Fourier transform of all GD 244 data in Figure~\ref{GD244ft}.  We used the prewhitening technique to identify pulsation modes and combination frequencies in GD 244, as with GD 66.  We also used prewhitening to reveal the doublet structure in the two highest amplitude pulsation modes.  As with GD 66, we do not expect that we have consistently measured the correct frequencies of the modes in these doublets in the presence of the contaminating window function.  However, the amplitudes are useful in estimating the inclination.
\clearpage

\begin{figure}
\caption{Fourier transform of GD 244.  This FT includes all individual nights of data included in Table~\ref{j_obsGD}.  We indicate the four pulsation modes and all combination frequencies that we reference in Table~\ref{tbl-GD244}.  F1+F3 is likely a blend with guide error.}
\label{GD244ft}
\end{figure}
\clearpage
In Figure~\ref{GD244_256ft} we show the deconstruction of the 256 second pulsation mode (F2) by prewhitening.  The Fourier transform (FT) of the reduced data in the region of F2 is shown in Figure~\ref{GD244_256ft}a.  The window function for the GD 244 data set is in Figure~\ref{GD244_256ft}b.  Figure~\ref{GD244_256ft}c shows the FT of the lightcurve with the largest peak removed.  Panel c is a prewhitened FT that reveals additional low amplitude signal which was previously hidden in the window function of the highest amplitude peak.  Figure~\ref{GD244_256ft}d is an FT of a lightcurve with the two highest amplitude peaks fitted and removed from the original lightcurve.  We were unable to fit any other statistically significant signals from the FT in the bottom panel.
\clearpage
\begin{figure}
\caption{ Deconstruction of F2 in GD 244.  a: A Fourier transform (FT) of GD 244 near 256 s (F2).  b: A window function obtained by taking an FT of a single sinusoid (with the same period and amplitude of the 256.56 s peak) that has been sampled in the same manner as the data.  c: An FT near F2 with a period of 256.56 s removed.  d: An FT near F2 with periods of 256.56 and 256.20 s removed.  Fitting and removing further peaks did not reduce the noise level. }
\label{GD244_256ft}

\end{figure}
\clearpage
A similar procedure showed that F1 is also consistent with a doublet.  F3, however, is consistent with a single frequency, although any putative companion peak would be in the noise if it scaled as the companions of F1 and F2.  F4 is not formally significant, with a false alarm probability near 1 in the whole data set,  but it appears above the noise at the same frequency in all three of the single month Fourier transforms so we have included it in Table~\ref{tbl-GD244}.  It does not show any combination peaks and therefore does not enter our analysis.  The complete list of pulsation modes is listed in Table~\ref{tbl-GD244} and presented visually in Figure~\ref{gd244periodogram}.  We did not detect the 294.6 s pulsation mode found by \citet{fon01} in GD 244.  All other pulsation modes that we list, except F4, were identified by \citet{fon01}.  There are no formally significant peaks above the noise level in the low frequency region of their FT. 
\clearpage
\begin{deluxetable}{rlllccrrr}
\tabletypesize{\scriptsize}
\tablewidth{0pt}
\tablecaption{GD 244 Periods and Mode Identifications \label{tbl-GD244}}
\tablehead{
\colhead{Mode Label} & \colhead{Frequency} & \colhead{Period} &\colhead{$\sigma_p$} & \colhead{Amplitude} &\colhead{$\sigma_{amp}$} & \colhead{$\Delta f$ \tablenotemark{a}} & \colhead{$\ell$} & \colhead{$m$ \tablenotemark{b}} \\
\colhead{} & \colhead{($\mu Hz$)} & \colhead{(sec)} & \colhead{(sec)} & \colhead{(mma)} & \colhead{(mma)} & \colhead{($\mu Hz$)} & \colhead{} & \colhead{}} 
\startdata
F1 &3255.919 &307.1329 &0.0001 &20.18 &0.17 &\nodata &1 &-1 \\
\nodata &3261.886 &306.5712 &0.0002 &5.02 &0.17 &5.966? &1 &+1 \\
F2 &3897.733 &256.5594 &0.0001 &12.31 &0.20 &\nodata &1? &-1? \\
\nodata &3903.255 &256.1964 &0.0001 &6.73 &0.20 &5.522? &1? &+1? \\
F3 &4926.697 &202.9758 &0.0001 &4.04 &0.21 &\nodata &1 &-1? \\
F4 &1103.656 &906.0795 &0.0056 &1.72 &0.21 &\nodata &$\leq 3$ &? \\
2F1$^-$ &6511.422 &153.5763 &0.0001 &2.25 &0.23 &0.417 &\nodata &\nodata \\
F1$^-$+F1$^+$ &6516.119 &153.4656 &0.0003 &0.95 &0.23 &1.686 &\nodata &\nodata \\
F1$^-$+F2$^-$ &7153.211 &139.7974 &0.0001 &2.30 &0.21 &0.442 &\nodata &\nodata \\
F1$^-$+F3$^-$\tablenotemark{c} &8182.615 &122.2103 &0.0002 &0.96 &0.21 &0.001 &\nodata &\nodata \\
2F2$^-$ &7795.472 &128.2796 &0.0001 &2.44 &0.21 &-0.006 &\nodata &\nodata \\
F2$^-$+F3$^-$ &8824.598 &113.3196 &0.0003 &0.53 &0.21 &-0.168 &\nodata &\nodata \\
\enddata
\tablenotetext{a}{For real pulsation modes, $\Delta f$ is the separation between the modes in the doublets.  For combination frequencies, $\Delta f$ is the frequency difference between the calculated and observed combination frequency (i.e., $\Delta f =F1+F2-[F1+F2]$).}
\tablenotetext{b}{The $m$ identifications in the table are based on frequency splitting alone, not on the size of the combination peaks.}
\tablenotetext{c}{The period of F1$^-$+F3$^-$ in GD 244 is close to the period of F6 in GD 66 (see Table~\ref{tbl-GD66}), so it is probable that the amplitude and frequency of this combination frequency are partially contaminated by the guide error of the telescope.}
\end{deluxetable}
\clearpage
\begin{figure}
\includegraphics[scale=.7,angle=-90]{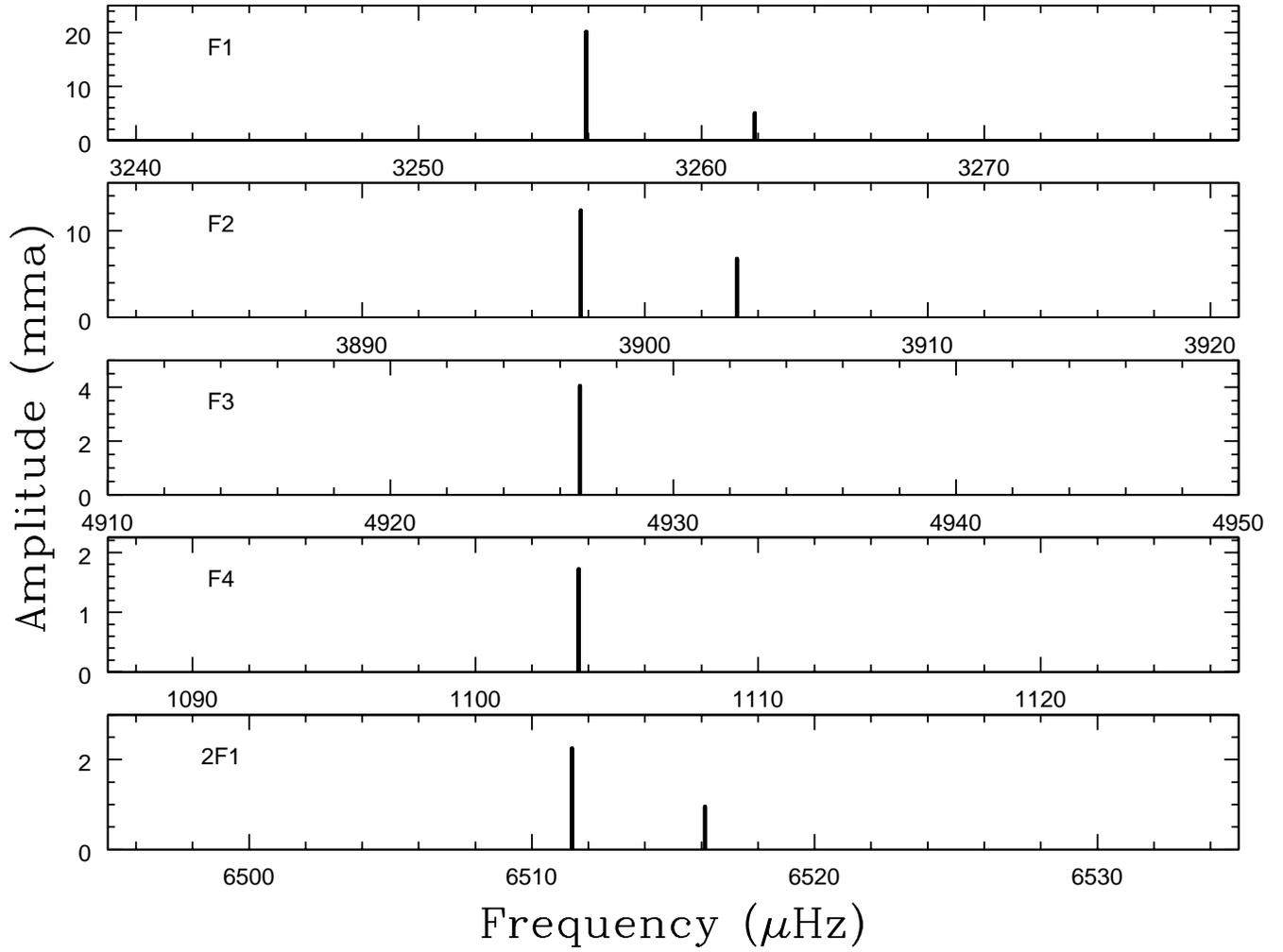}
\caption{ Prewhitened peaks in the GD 244 Fourier transform.  In addition to the parent modes, we include one combination frequency (2F1) in which we resolved fine structure. }
\label{gd244periodogram}
\end{figure}
\clearpage
In addition to the four modes detected, we identified six combination frequencies that were consistently present throughout the observing run.  In all cases, the highest amplitude peak among the combination frequency and its aliases fell within the error of the expected frequency.  We list the frequencies of the highest amplitude peaks in Table~\ref{tbl-GD244}.  The frequency errors were sufficiently small that there was no confusion between the harmonic of the largest component (presumed $m=-1$) and the sum of the $m=-1$ and $m=+1$ components.

We use the doublet structure at 307 seconds (F1) to estimate the inclination of the pulsation axis of GD 244.  We assume that the doublet structure in these two pulsation modes results from viewing $\ell = 1$ modes at high inclination so that the third (central, $m=0$) mode does not appear.  As discussed previously, our results do not depend sensitively on this assumption.  Using the maximum amplitude in the prewhitened FT of F1 as an estimate for the amplitude of the $m = 0$ peak, we apply the \citet{pes85} method and find a minimum possible inclination of $80$ degrees.  Recall that Figure~\ref{G1111gg} shows that at high inclination, the amplitudes of combination frequencies with $\ell=1$ and same-$m$ parent modes have no dependence on inclination.

\begin{figure}
\includegraphics[scale=.7,angle=-90]{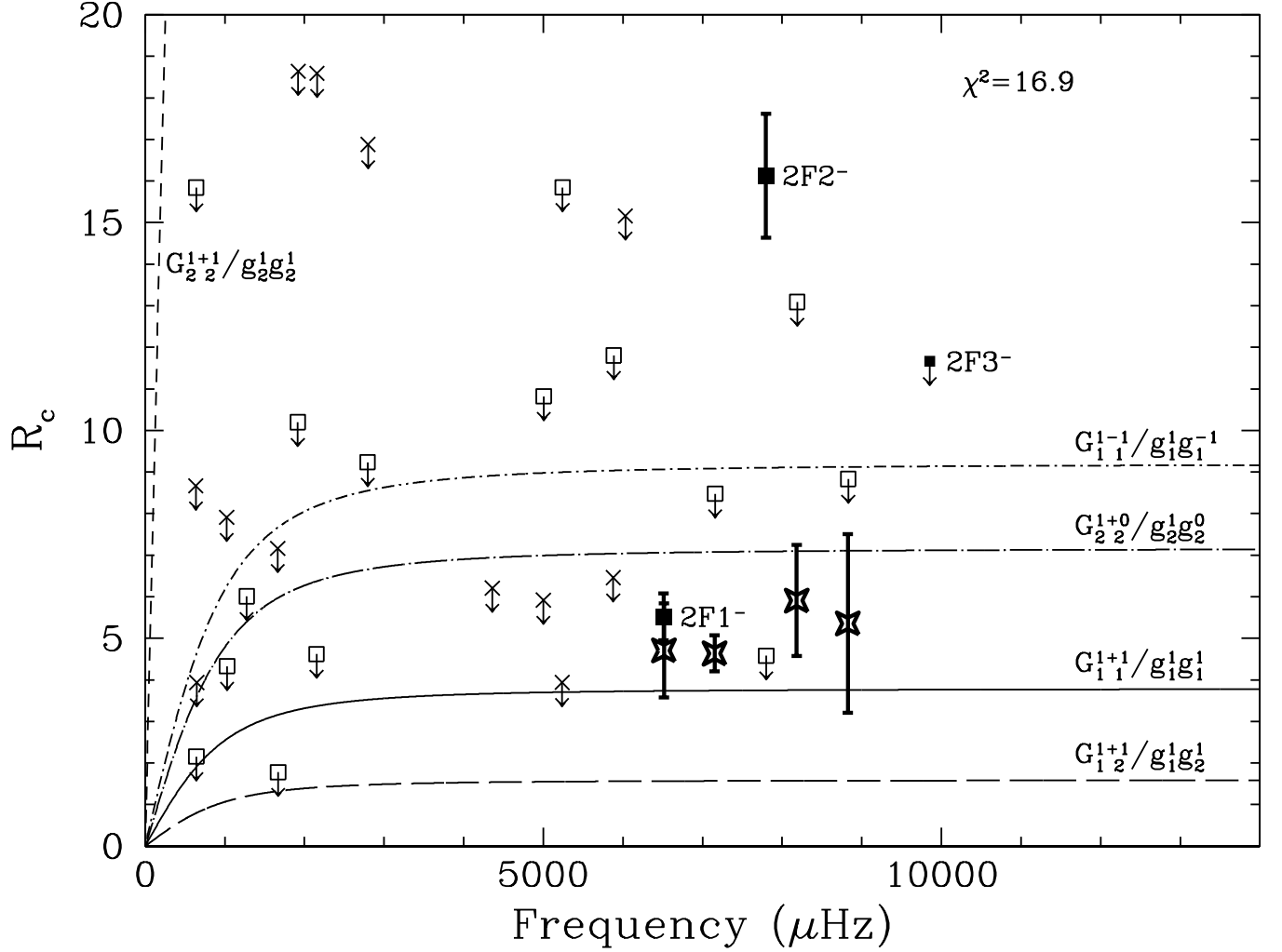}
\caption{ Ratio of combination to parent mode amplitudes ($R_c$) for GD 244.  The lines are theoretical predictions for $G_{1~2}^{1+1} \diagup g_1^1 g_2^1 (\Theta_\circ=80^\circ)$ (long-dashed line), $G_{1~1}^{1+1} \diagup g_1^1 g_1^1 (\Theta_\circ=80^\circ)$ (solid line), $G_{2~2}^{1+0} \diagup g_2^1 g_2^0 (\Theta_\circ=80^\circ)$ (dot-long-dashed line), $G_{1~1}^{1-1} \diagup g_1^1 g_1^{-1} (\Theta_\circ=80^\circ)$ (dot-dashed line), and $G_{2~2}^{1+1} \diagup g_2^1 g_2^1 (\Theta_\circ=80^\circ)$ (dashed line).  The data points are the detected harmonics or limits (filled squares), detected cross combination frequencies (stars), limits for the same-$m$ cross combinations (crosses), and limits for the different-$m$ cross combinations (open squares).  The downward arrows on the limits indicate that the points represent maximum values. }
\label{RcGD244}
\end{figure}

Figure~\ref{RcGD244} is a plot of the theoretical predictions for GD 244 ($R_c$ with $\Theta_\circ=80^\circ$) and the observed amplitudes of the six detected combination frequencies for comparison.  We also include the observed noise limit (indicated by crosses for the same-$m$ combinations and open squares for the different-$m$ combinations) in cases where there was no combination frequency detected.  The downward arrows imply an upper limit.  When we apply the GD 66 calibration of $2\beta+\gamma=-9.35$ to GD 244 (in Figure~\ref{RcGD244}), the predictions for the $\ell=1,1$ line resemble the observations for our detected combination frequencies except for the $R_c$ of 2F2.  The reduced $\chi^2$ is 16.9 if all the modes are assumed to be $\ell=1$, but 9.2 if we leave out the combinations of F2.  Obviously we could reduce $\chi^2$ further if we varied $2\beta+\gamma$ as before, but part of our exercise is to establish that we can conduct mode identification without fitting parameters, so that it is possible to measure the $\ell$ of hot ZZ Cetis with only a single detected combination frequency.

Turning to the harmonics, which are known to be same-$\ell$ and therefore to have well-separated predictions for $R_c$, 2F1 and the limit for 2F3 demand that F1 and F3 be $\ell=1$ modes.  The high measurement for 2F2 suggests that $\ell$ may not be 1.  However, if we assume $\ell_{F2}=2$, then $\chi^2 > 400$ because $R_c$ for a harmonic of an $\ell=2$, $m=1$ mode is predicted to be near 90.  Even if we relax the assumptions used to calculate inclination and adjust it to minimize $\chi^2$ for the assumption of $\ell_{F2}=2$, then at $\Theta_\circ=67^\circ$, $\chi^2=17.3$.  So even under the best assumptions, the identification of F2 as $\ell=2$ would yield a worse fit than if we let it be $\ell=1$.  Moreover, its similar frequency splitting to F1 suggests that it is $\ell=1$, but without better sampling with WET this is not a secure statement.  Therefore, we have left question marks next to the identification of F2 in Table~\ref{tbl-GD244}.  It is possible that the harmonic of F2 is inflated by some independent unresolved pulsation mode, but our data are insufficient to test this possibility.  Finally, for the smallest mode, F4, the limit on the harmonic constrains it to be $\ell \leq 3$, which is not useful.

Because F1 and F2 are multiplets, we expect fine structure in their harmonics and combinations, and indeed we observe secondary peaks near both harmonics at the sum of the presumed $m=-1$ and $m=1$ parent modes.  Unfortunately, the theoretical lines for these cross terms in $m$ are not well-separated from similar cross terms in an $\ell=2$ mode (see Figure~\ref{RcGD244}).  Likewise, the F1+F2 peak shows multiplet structure, but we are unable to reliably dissect it into individual modes.  The limits we have measured for the cross terms of different-$m$s are all consistent with the $\ell=1$ identifications, but would not be sufficient alone to reach that conclusion.
\clearpage
\clearpage
\subsubsubsection{\it{G117-B15A}}
\label{G117section}

G117-B15A is one of the hottest known ZZ Ceti stars.  \citet{mr76} confirmed the star's variability and \citet{kep82} found six pulsation modes.  The dominant mode, at 215 seconds, is stable in amplitude and phase such that G117-B15A is the most precise optical clock known \citep{kep00a}.  We use published measurements for G117-B15A obtained from the WET campaign XCov 4 in 1990 May \citep{kep95b}.  They list the highest amplitude modes as $\ell=1$. 
\clearpage
\begin{deluxetable}{rlllccrrrc}
\tablecolumns{10}
\tabletypesize{\scriptsize}
\tablewidth{0pt}

\tablecaption{Periods and Mode Identifications for Published Data with Combination Frequencies \label{tbl-published}}
\tablehead{
\colhead{Mode Label} & \colhead{Frequency} & \colhead{Period} & \colhead{$\sigma_p$} & \colhead{Amplitude} & \colhead{$\sigma_{amp}$} & \colhead{$\Delta f$ \tablenotemark{a}} & \colhead{$\ell$} & \colhead{$m$} & \colhead{Reference} \\
\colhead{} & \colhead{($\mu Hz$)} & \colhead{(sec)} & \colhead{(sec)} & \colhead{(mma)} & \colhead{(mma)} &\colhead{($\mu Hz$)} & \colhead{} & \colhead{} & \colhead{} \\
\cline{1-10} \\
\multicolumn{10}{c}{G117-B15A}} 
\startdata
F1 &4646.909 &215.1968 &0.0007 &19.15 &0.39 &\nodata &1 &0 &1 \\
F2 &3288.91 &304.052 &0.004 &6.89 &0.44 &\nodata &1 &0 &1 \\
F3 &3697.47 &270.455 &0.004 &5.47 &0.45 &\nodata &1? &0 &1 \\
2F1 &9293.68 &107.600 &0.004 &1.06 &0.45 &0.14 &\nodata &\nodata &1 \\
F1+F2 &7956.59 &125.682 &0.006 &0.90 &0.45 &-20.77 &\nodata &\nodata &1 \\
F1+F3 &8344.74 &119.836 &0.003 &1.60 &0.45 &-0.36 &\nodata &\nodata &1 \\
F1-F2	&1357.6 &736.60 &\nodata &0.90 &\nodata &0.4 &\nodata &\nodata &1 \\
\cutinhead{G185-32}
F1 &4635.3 &215.74 &\nodata &1.93 &0.07 &\nodata &1 or 2 &0 &2 \\
F2 &2701.2 &370.21 &\nodata &1.62 &0.07 &\nodata &1 or 2 &0 &2 \\
F3 &7048.8 &141.87 &\nodata &1.43 &0.07 &\nodata &3 &0 &2 \\
F4 &3317.8 &301.41 &\nodata &1.13 &0.07 &\nodata &1 or 2 &0 &2 \\
F5 &3335.6 &299.79 &\nodata &0.95 &0.07 &\nodata &1 or 2 &0 &2 \\
F6 &13784.9 &72.54 &\nodata &0.93 &0.07 &\nodata &1 or 2 &0 &2 \\
2F3 &14097.7 &70.93 &\nodata &0.69 &0.07 &-0.1 &\nodata &\nodata &2 \\
F3-F6 &6736.1 &148.45 &\nodata &0.57 &0.07 &0.0 &\nodata &\nodata &2 \\
\enddata
\tablecomments{We have not included a complete list of eigenmodes for each star.  Instead, we have only included eigenmodes relevant to this study.  The $\ell$ and $m$ identifications are from our analysis.}
\tablenotetext{a}{$\Delta f$ is the frequency difference between the calculated and observed combination frequency (i.e., $\Delta f =F1+F2-[F1+F2]$).}
\tablerefs{(1) \citet{kep95b}; (2) \citet{cas04}.}
\end{deluxetable}
\clearpage
The 215 s pulsation mode of G117-B15A has a large central peak and two possible adjacent peaks regarded by \citet{kep95b} as ``low probability'' because they do not rise very far above the already low noise level.  We have used the size of the adjacent peaks to constrain the inclination of this star, under the assumption that the central peak is an $\ell=1$, $m=0$ mode.  We find that the star has a maximum inclination of five degrees (nearly pole-on), and this result is not sensitive to the assumption of $\ell=1$.  In order to appear as a singlet, modes of any $\ell$ must be viewed at low inclination.
\clearpage
\begin{figure}
\includegraphics[scale=.7,angle=-90]{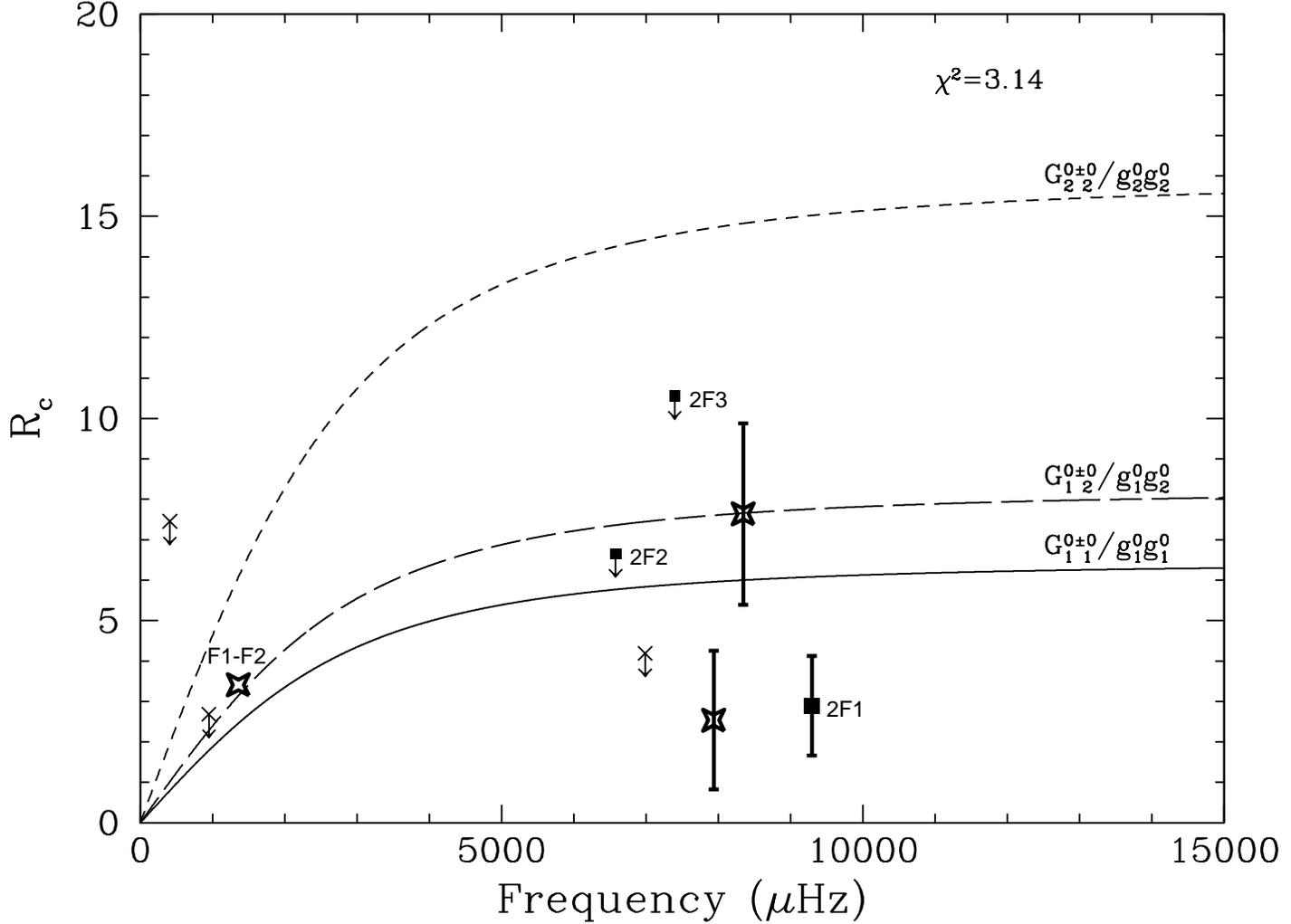}
\caption{Ratio of combination to parent mode amplitudes ($R_c$) for G117-B15A.  The lines are theoretical predictions for $G_{1~1}^{0+0} \diagup g_1^0 g_1^0 (\Theta_\circ=5^\circ)$ (solid line), $G_{1~2}^{0+0} \diagup g_1^0 g_2^0 (\Theta_\circ=5^\circ)$ (long-dashed line), and $G_{2~2}^{0+0} \diagup g_2^0 g_2^0 (\Theta_\circ=5^\circ)$ (dashed line).  The data points are the detected harmonics or limits (filled squares), detected cross combination frequencies (stars), and limits for the cross combinations (crosses).  The downward arrows on the limits indicate that the points represent maximum values.  There are no error bars for F1-F2 because there were none reported by \citet{kep95b}. }
\label{RcG117B15A}
\end{figure}
\clearpage
Figure~\ref{RcG117B15A} is a plot of the theoretical predictions for G117-B15A ($R_c$ with $\Theta_\circ=5^\circ$) and the observed amplitudes of the four detected combination frequencies for comparison.  We also include the observed noise limit (indicated by the crosses) in cases where there was no combination frequency detected.  The downward arrows below the crosses indicate an upper limit.  Just as with GD 66, Wu's theory predicts high amplitudes for all of the combination frequencies that we detect and low amplitudes for those we do not detect.  When we apply the GD 66 calibration of $2\beta+\gamma=-9.35$ to G117-B15A (in Figure~\ref{RcG117B15A}), the predictions are consistent with the observations for our detected combination frequencies, though the scatter is considerably worse than in GD 66.  The reduced $\chi^2$ is 3.14 if all the modes are assumed to be $\ell=1$.  The combination F1-F2, located in the lower left quadrant of Figure~\ref{RcG117B15A}, is not included in this $\chi^2$, because no formal errors were reported by \citet{kep95b}.  If we assume the error for the amplitude of this combination is greater than the largest errors in the plot, then it gives no information about the $\ell$ of its parents, but is consistent with the low frequency roll-off of Wu's theory.

\citet{rob95} established that $\ell=1$ for F1 by comparing time-resolved spectroscopy observations in the UV and optical wavelengths.  The amplitude of the harmonic of this mode, combined with our analysis, confirms this $\ell=1$ identification; the observed $R_c$ for the harmonic of F1 is too small to be consistent with that for $\ell_i=\ell_j=2$, which is 2.5 times greater than $R_c$ for $\ell_i=\ell_j=1$ and 5 greater than the $R_c$ we observe.  Likewise, the limit for the harmonic of F2 is sufficiently low to constrain this mode to be $\ell=1$ with a 98.7 percent confidence level.  The limit for 2F3 is more ambiguous, but clearly requires that $\ell \leq 2$ for F3.  We list our mode identifications in Table~\ref{tbl-published}.  These results are consistent with the seismological analyses of both \citet{brad98} and \citet{brass93}, where the choice of $\ell=1$ for these three peaks yielded reasonable physical parameters.

\subsubsubsection{\it{G185-32}}

\citet{mc81} discovered the DA white dwarf G185-32 to be a relatively low amplitude multi-periodic ZZ Ceti star on the basis of its $(G - R)$ colors \citep[see][]{gre76}.  The largest amplitude peaks have periods of 71, 73, 142, 216, 301, and 370 seconds.  G185-32 is unique among the ZZ Cetis in that it has a harmonic (at 71 s) that is sometimes measured to be larger than its parent frequency (at 142 s).  This has led to disagreement over whether the harmonic is a pulse shape artifact, or whether there are resonances between real eigenmodes that happen to be harmonically related \citep{cas04}.  Most recently, \citet{thomp04} has found that identifying the 142 s mode as high $\ell$ ($\ell=4$) can explain all of the available observations, which include time-resolved UV spectroscopy from HST \citep{kep00b}, time-resolved optical spectroscopy from Keck \citep{thomp04}, and time-series photometry from WET \citep{cas04}.  An $\ell=4$ mode cancels itself in the integration over the visible hemisphere for almost any value of inclination, but its harmonic may not, because it has a surface distribution with characteristics of lower $\ell$.
This allows a harmonic to appear larger than its fundamental.  In our analysis, we apply the theory of Wu under the assumption that the peak at 142 s is the parent mode of a harmonic at 71 s.  We use published measurements for G185-32 obtained from the WET campaign XCov 8 in 1992 September \citep{cas04}.

With the exception of the 142 s mode, \citet{kep00b} identify all the other modes as either $\ell=1$ or $\ell=2$ based on HST data alone.  Using independent temperature constraints they choose $\ell=1$ for all of these modes.  Under this identification, F1\footnote{Our nomenclature (see Table~\ref{tbl-published}) does not follow \citet{thomp04}, \citet{cas04}, nor \citet{kep00b}.  We have labeled the modes in order of highest amplitudes from the WET data.}, at 215 s, is the highest amplitude $\ell=1$ mode, and we use it to estimate the inclination.  \citet{cas04} detect only one peak at 215 s, so we presume it to be the $m=0$ component and use the neighboring noise limit to approximate the size of the $m=\pm 1$ peaks.  The \citet{pes85} method then yields $\Theta_\circ = 13$ degrees.  If the inclination were greater than this, we would see the $m=\pm 1$ members of the multiplet above the noise.  Our results are robust even if the $\ell$ identification is not; if F1 is actually an $\ell=2$ mode, then the inclination of the star is less than 7 degrees, and still in the range where our results are insensitive to the inclination (see Figure~\ref{Ggg}).  
\clearpage
\begin{figure}
\includegraphics[scale=.7,angle=-90]{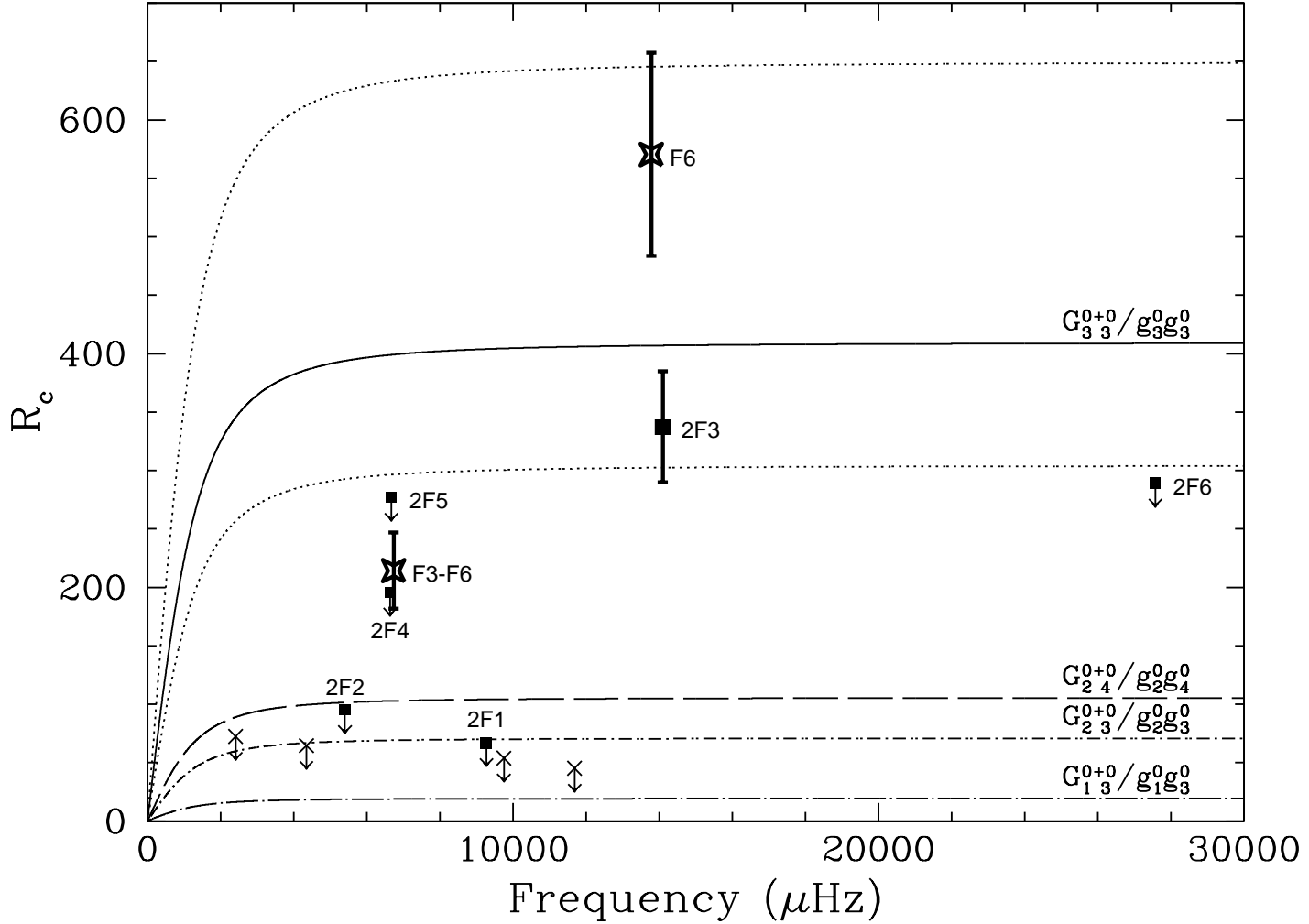}
\caption{Ratio of combination to parent mode amplitudes ($R_c$) for G185-32.  The lines are theoretical predictions for $G_{3~3}^{0+0} \diagup g_3^0 g_3^0(\Theta_\circ=13^\circ)$ (solid line), $G_{2~4}^{0+0} \diagup g_2^0 g_4^0(\Theta_\circ=13^\circ)$ (long-dashed line), $G_{2~3}^{0+0} \diagup g_2^0 g_3^0(\Theta_\circ=13^\circ)$ (dot-dashed line), and $G_{1~3}^{0+0} \diagup g_1^0 g_3^0(\Theta_\circ=13^\circ)$ (dot-long-dashed line).  The data points are the detected harmonics or limits (filled squares), detected cross combination frequencies (stars), and limits for the cross combinations involving the F3 pulsation mode only (crosses).  The downward arrows on the limits indicate that the points represent maximum values.  The dotted lines on either side of the $\ell_i = \ell_j = 3$ line are the $R_c$ predictions for extremum in inclination: $\Theta_\circ=20^\circ$ (top) and $\Theta_\circ=0^\circ$ (bottom).  Wu's predictions for $G_{4~4}^{0+0} \diagup g_4^0 g_4^0(\Theta_\circ=13^\circ)$ are not shown because they reach values above $R_c \sim 3600$. } 
\label{RcG18532}
\end{figure}
\clearpage
Figure~\ref{RcG18532} is a plot of the theoretically predicted combination frequency amplitude ratios for G185-32 ($R_c$ with $\Theta_\circ=13^\circ$) and the observed amplitudes of the detected combination frequencies for comparison.  For clarity, the $\ell=1,1$ and $\ell=2,2$ lines are not included in Figure~\ref{RcG18532}, but fall below the lowest line shown ($\ell=1,3$).  Unlike the harmonics found in GD 66, GD 244, and G117-B15A, $R_c$ for the harmonic of F3 in G185-32 is more than 50 times greater than the prediction for $\ell_i=\ell_j=1$ and more than 15 times the prediction for $\ell_i=\ell_j=2$.  The best explanation for this is that F3 is a high $\ell$ mode, just as \citet{thomp04} claim for independent reasons.  The value of $\ell$ that best explains our $R_c$ is $\ell=3$ rather than $\ell=4$.  However, in the case of high $\ell$, our theory suffers from two problems.  First, the geometric factor in Wu's theory varies more rapidly with inclination for $\ell =3$ and $4$ (instead of the gradual changes for low $\ell$ shown in Figure~\ref{Ggg}, see Figure~\ref{G34} in Appendix A).  We have accommodated this by including a range of inclinations about the 13 degree nominal value.  We show this range as dotted lines in Figure~\ref{RcG18532}, which indicate how $R_c$ for $\ell_i = \ell_j = 3$ changes between $\Theta_\circ = 0^\circ$ and $20^\circ$.  $R_c$ for 2F3 falls within the prediction for $\ell_{F3}=3$.  The second problem is our use of a constant bolometric correction that is really only appropriate for $\ell=1$.  This is a limitation inherent to the analytic theory as Wu presented it, and could be addressed by employing numerical models, but this would be not be consistent with our objective of having a quick and easily applicable method for mode identification in large numbers of ZZ Ceti stars.  We have calculated bolometric corrections for higher $\ell$, and the differences are not large enough to change the $\ell=3$ identification for F3.  So the inconsistency of our $\ell=3$ identification and the $\ell=4$ identification of \citet{thomp04} remains a mystery.  Fortunately, for seismological work it is useful to identify a mode as high $\ell$, even if the exact value is unknown.  The density of modes in the models at high $\ell$ is so large that they do not contribute seismological constraints because the radial overtone ($n$ or $k$) is unknown, but misinterpreting a high $\ell$ mode as $\ell = 1$ or 2 would lead the models astray.  So even if a mode identification method based on measurements of $R_c$ alone cannot identify high $\ell$ modes precisely, it is still useful for its ability to discriminate between modes of high and low $\ell$.

In addition to the identification of F3 as high $\ell$, the limits on $R_c$ for the harmonics of F1, F2, F4, F5, and F6 constrain that $\ell \leq 2$ for these modes, consistent with the tentative mode identifications of \citet{cas04}.  Our results are also consistent with the seismological analysis of \citet{brad05}, including his high $\ell$ identification for F3, with the exception of his identification of F6 as high $\ell$.  Unfortunately, we are not able to definitively assign any $\ell$ values to these modes due to their intrinsically low amplitudes \citep[see][]{clemens94,thomp04}.  Likewise, the only detected cross term, which is a combination with the high $\ell$ mode F3, is not able to narrow the choice of $\ell$.  Indeed, we are not even able to determine whether the mode we call F6 is a combination with F3 and a mode at 148 s or whether the latter is F3-F6.  We have used the latter identification in Table~\ref{tbl-published}, following \citet{cas04} and consistent with \citet{brad05}.

\subsubsection{Stars Without Detected Combination Frequencies}
\label{nodetection}

\subsubsubsection{\it{L19-2}}

\citet{mc77} discovered the variability of L19-2, a hot, low-amplitude ZZ Ceti star.  \citet{ow82} presented a comprehensive analysis of single site data, and were able to assign tentative values of $\ell$ to the pulsation modes.  \citet{brad01} revised these identifications in light of theoretical improvements, finding three modes of $\ell=1$ and two of $\ell=2$.  L19-2 was the subject of the WET campaign XCov 12 in 1995 April, on which \citet{sul95} presented a preliminary paper.  We have re-reduced and analyzed the archival WET data to search for combination frequencies, and found none detectable above the noise limit of the Fourier transform.

\begin{figure}
\includegraphics[scale=.7,angle=-90]{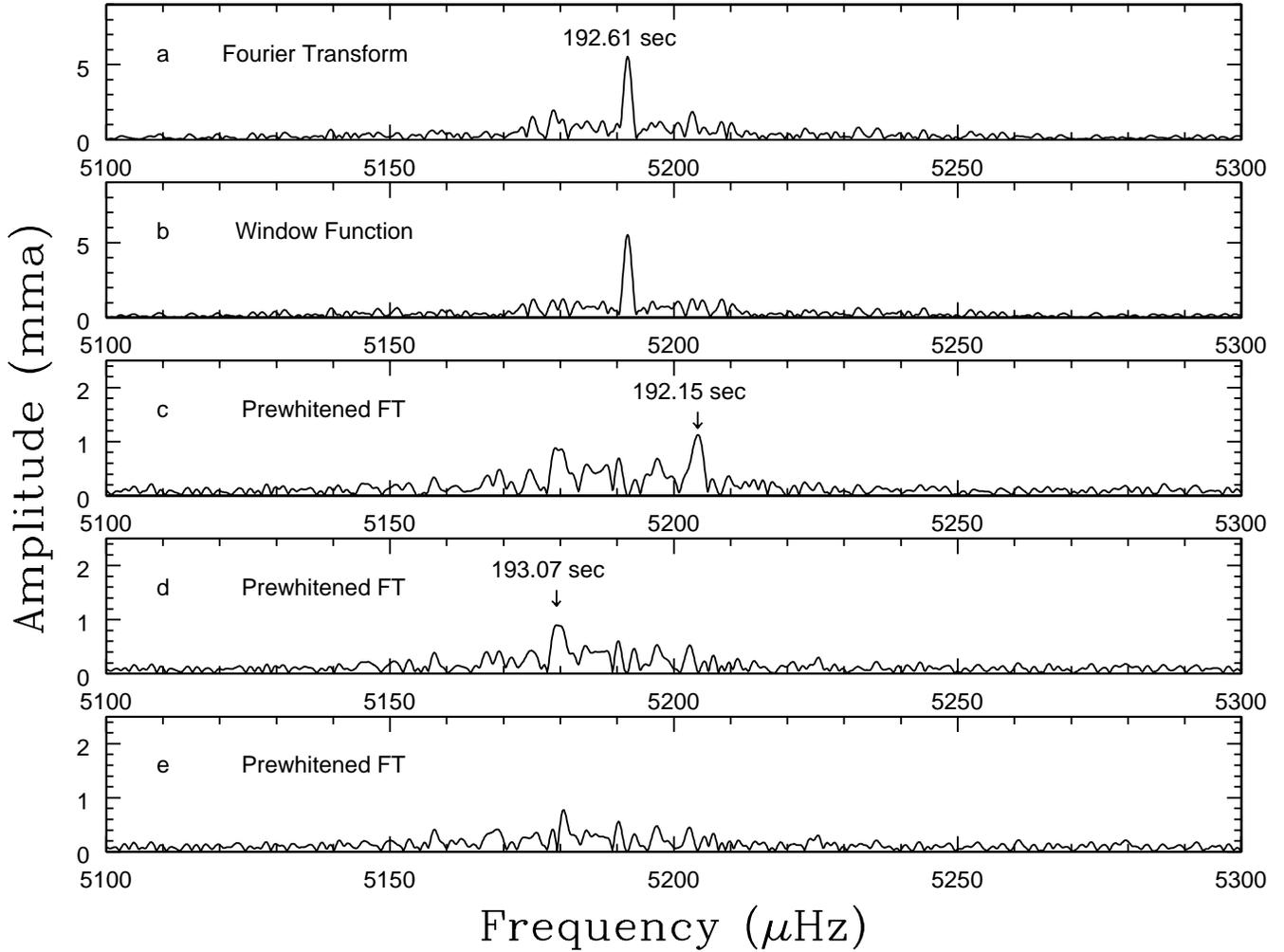}
\caption{ Deconstruction of F1 in L19-2.  a: A Fourier transform of L19-2 near 192 s (F1).  b: A window function obtained by taking an FT of a single sinusoid (with the same period and amplitude of the 192.61 s peak) that has been sampled in the same manner as the data.  c: An FT near F1 with a period of 192.61 s removed.  d: An FT near F1 with periods of 192.61 and 192.15 s removed.  e: An FT near F1 with periods of 192.61, 192.15, and 193.08 s removed.  }
\label{L192_192ft}
\end{figure}

\begin{figure}
\includegraphics[scale=.7,angle=-90]{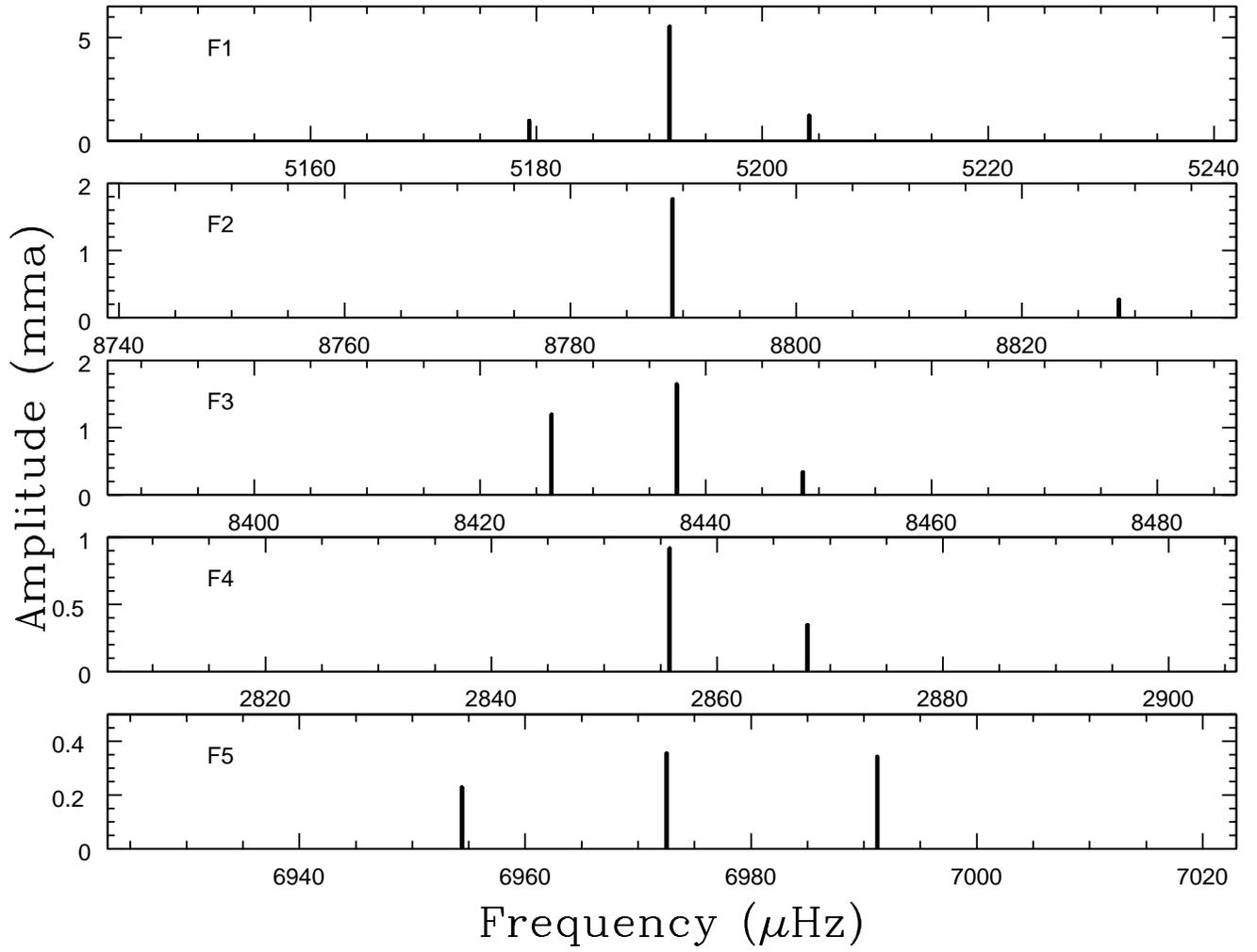}
\caption{Prewhitened peaks in the L19-2 Fourier transform. }
\label{l192periodogram}
\end{figure}

As with GD 66, we used the prewhitening technique to measure pulsation frequencies in L19-2, and to reveal the multiplet structure in the highest amplitude modes.  In Figure~\ref{L192_192ft} we show the 192 s pulsation mode (F1) as a typical example.  Figure~\ref{L192_192ft}a is the original FT in the region of F1.  Figure~\ref{L192_192ft}b is the window function of the 192.61 second peak.  The remaining panels show the results of our iterative fitting and removal of three pulsation frequencies.  Note that there appears to be a significant residual near the $m=-1$ component.  This is mysterious, but not unprecedented \citep{kaw95}.  The identified periods of L19-2 are listed in Table~\ref{tbl-l192} and presented visually in Figure~\ref{l192periodogram}.  We do not detect any combination frequencies in L19-2.  However, the noise level is sufficiently low, that the non-existence of combinations, particularly harmonics, constrains the $\ell$ values of the modes present.

\clearpage
\clearpage
\begin{deluxetable}{rlllccrrr}
\tablecolumns{9}
\tabletypesize{\scriptsize}
\tablewidth{0pt}
\tablecaption{L19-2 Periods and Mode Identifications \label{tbl-l192}}
\tablehead{
\colhead{Mode Label} & \colhead{Frequency} & \colhead{Period} & \colhead{$\sigma_p$} & \colhead{Amplitude} & \colhead{$\sigma_{amp}$} & \colhead{$\Delta f$ \tablenotemark{a}} & \colhead{$\ell$} & \colhead{$m$ \tablenotemark{b}} \\
\colhead{} & \colhead{($\mu Hz$)} & \colhead{(sec)} & \colhead{(sec)} & \colhead{(mma)} & \colhead{(mma)} &\colhead{($\mu Hz$)} & \colhead{} & \colhead{}} 
\startdata
F1 &5179.362 &193.0740 &0.0019 &0.973 &0.064 &-12.433 &1 &-1 \\
\nodata &5191.795 &192.6116 &0.0004 &5.535 &0.065 &\nodata &1 &0 \\
\nodata &5204.160 &192.1540 &0.0015 &1.216 &0.063 &12.365 &1 &+1 \\
F2 &8789.054 &113.7779 &0.0004 &1.766 &0.067 &\nodata &1 or 2 &0 \\
\nodata &8828.619 &113.2680 &0.0024 &0.271 &0.067 &39.565 &1 or 2 &+2 \\
F3 &8426.324 &118.6757 &0.0006 &1.191 &0.070 &-11.109 &1 or 2 &-1 \\
\nodata &8437.433 &118.5195 &0.0005 &1.641 &0.071 &\nodata &1 or 2 &0 \\
\nodata &8448.580 &118.3631 &0.0023 &0.339 &0.069 &11.147 &1 or 2 &+1 \\
F4 &2855.785 &350.1664 &0.0073 &0.918 &0.068 &\nodata &1 or 2 &0 \\
\nodata &2868.023 &348.6722 &0.0192 &0.347 &0.068 &12.238 &1 or 2 &+1 \\
F5 &6954.415 &143.7935 &0.0047 &0.228 &0.069 &-18.112 &$\leq 3$ &-1? \\
\nodata &6972.527 &143.4200 &0.0030 &0.354 &0.070 &\nodata &$\leq 3$ &0 \\
\nodata &6991.176 &143.0375 &0.0031 &0.341 &0.069 &18.649 &$\leq 3$ &+1 \\
\enddata
\tablenotetext{a}{$\Delta f$ is the separation between the modes in the multiplets and the $m=0$ member. }
\tablenotetext{b}{The $m$ identifications in the table are based on frequency splitting alone, not on the size of the combination peak limits.}
\end{deluxetable}
\clearpage

\begin{figure}
\includegraphics[scale=.7,angle=-90]{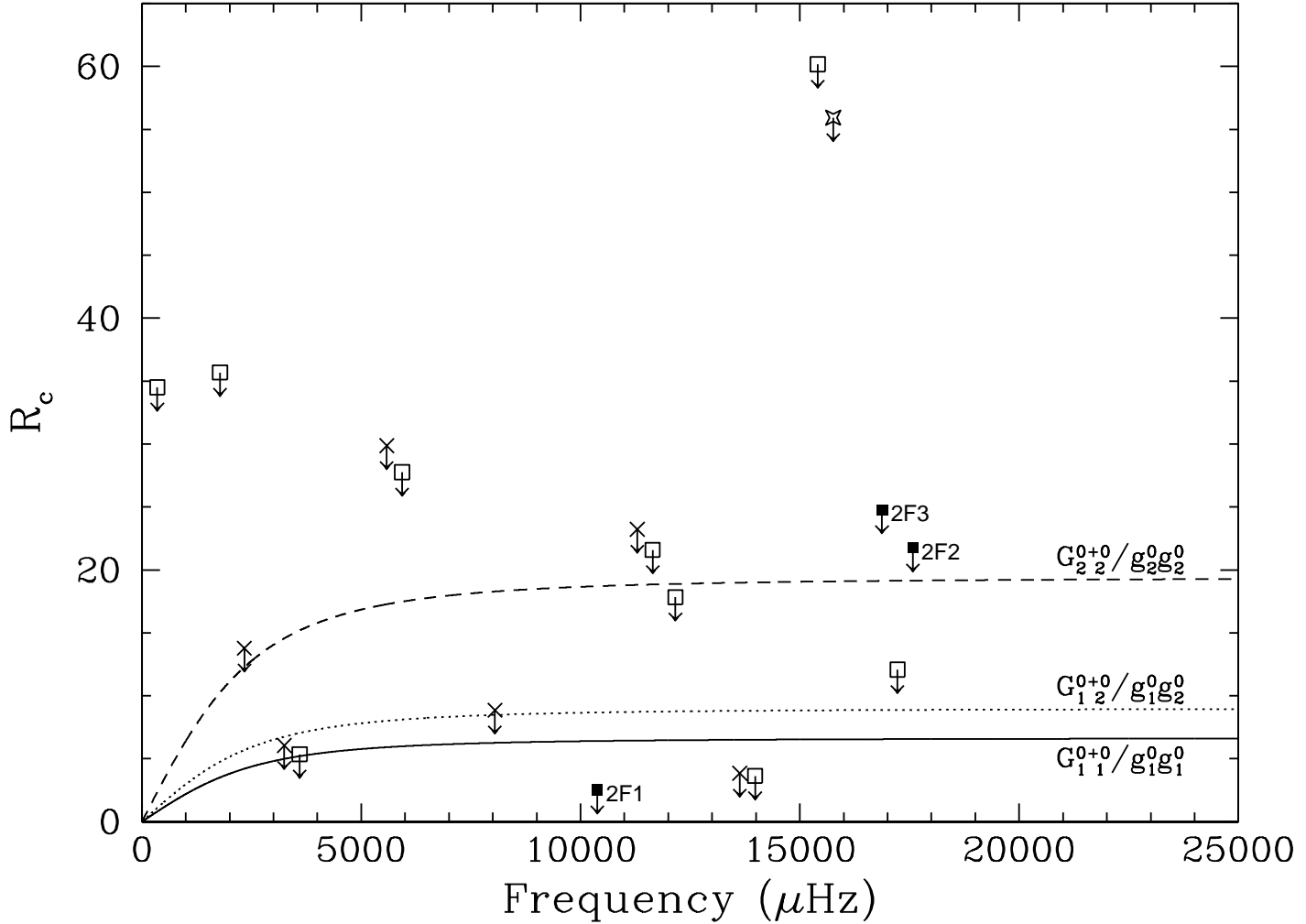}
\caption{Ratio of combination to parent mode amplitudes ($R_c$) for L19-2.  The lines are theoretical predictions for $G_{1~1}^{0+0} \diagup g_1^0 g_1^0 (\Theta_\circ=16^\circ)$ (solid line), $G_{1~2}^{0+0} \diagup g_1^0 g_2^0 (\Theta_\circ=16^\circ)$ (dotted line), and $G_{2~2}^{0+0} \diagup g_2^0 g_2^0 (\Theta_\circ=16^\circ)$ (dashed line).  The data points are the limits on the harmonics (filled squares) and limits for the cross combinations for the presumed \citep{brad01} $\ell=1,1$ combinations (crosses), $\ell=1,2$ combinations (open squares), and an $\ell=2,2$ combination (star).  The downward arrows on the limits indicate that the points represent maximum values. }
\label{RcL192}
\end{figure}
\clearpage
We have used the multiplet structure at 192 seconds (F1) in Figure~\ref{L192_192ft} to estimate the inclination of the pulsation axis of L19-2, finding an inclination of $16$ degrees.  In Figure~\ref{RcL192}, we have plotted the theoretical predictions for combination frequency amplitudes in L19-2 ($R_c$ with $\Theta_\circ=16^\circ$).  Instead of measured ratios $R_c$, we have included the observed noise limit (indicated by the filled squares, crosses, open squares, and stars) at the frequencies where we expect combinations to be detected.  The downward arrows indicate that all data points are limits on detections.  The limit for the harmonic of F1 implies that mode is $\ell=1$, which is consistent with how \citet{brad01} identified it based on its multiplet structure.  The limits for F2, F3, and F4 imply that they are $\ell \leq 2$, again consistent with the multiplet structure that suggests F2 is $\ell=2$ and F3 and F4 are $\ell=1$ \citep{brad01}.  The limit for 2F5 constrains F5 to be $\ell \leq 3$.

It is gratifying to find that non-detections of combination frequencies can provide useful seismological information, and that these corroborate independent methods in the case of L19-2.  We note that Wu's predictions suggest that the WET data are at the threshold of detecting the combination frequencies in L19-2.  Larger telescope data on this star might be useful as a further test of the reliability of Wu's theory for mode identification.

\subsubsubsection{\it{GD 165}}

The pulsation pattern of GD 165 is very similar to that of L19-2.  Both have their two primary pulsations near 120 and 193 seconds.  \citet{bm90} discovered GD 165 to be a ZZ Ceti star as predicted by temperatures acquired from spectroscopic analysis placing it within the theoretical ZZ Ceti instability strip. 

\begin{figure}
\includegraphics[scale=.7,angle=-90]{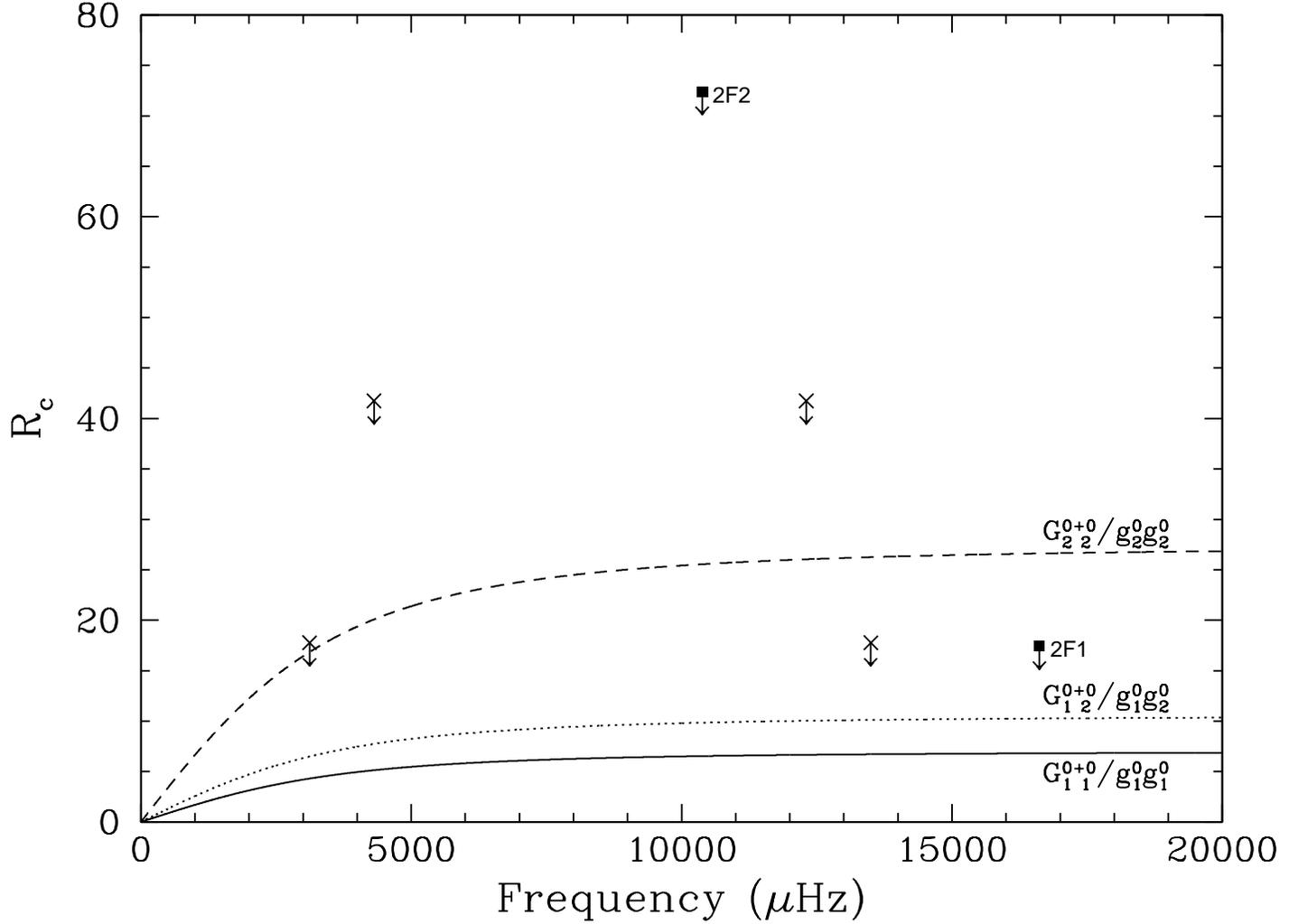}
\caption{Ratio of combination to parent mode amplitudes ($R_c$) for GD 165.  The lines are theoretical predictions for $G_{1~1}^{0+0} \diagup g_1^0 g_1^0 (\Theta_\circ=25^\circ)$ (solid line), $G_{1~2}^{0+0} \diagup g_1^0 g_2^0 (\Theta_\circ=25^\circ)$ (dotted line), and $G_{2~2}^{0+0} \diagup g_2^0 g_2^0 (\Theta_\circ=25^\circ)$ (dashed line).  The data points are the limits on the harmonics (filled squares) and limits for the cross combinations (crosses).  The downward arrows on the limits indicate that the points represent maximum values. }
\label{RcGD165}
\end{figure}

We included the data for GD 165 from WET observations (XCov 5, 1990 May) and from lightcurves obtained with the CFHT presented in an analysis by \citet{ber93}.  The two primary pulsations of GD 165 have multiplet structure, to which we applied the \citet{pes85} method and found 25 degrees for the inclination.
\clearpage
\begin{deluxetable}{rlllccrrrc}
\tablecolumns{10}
\tabletypesize{\scriptsize}
\tablewidth{0pt}
\tablecaption{Periods and Mode Identifications for Published Data without Combination Frequencies \label{tbl-pubnocomb}}
\tablehead{
\colhead{Mode Label} & \colhead{Frequency} & \colhead{Period} & \colhead{$\sigma_p$} & \colhead{Amplitude} & \colhead{$\sigma_{amp}$} & \colhead{$\Delta f$ \tablenotemark{a}} & \colhead{$\ell$} & \colhead{$m$ \tablenotemark{b}} & \colhead{Reference} \\
\colhead{} & \colhead{($\mu Hz$)} & \colhead{(sec)} & \colhead{(sec)} & \colhead{(mma)} & \colhead{(mma)} &\colhead{($\mu Hz$)} & \colhead{} & \colhead{} & \colhead{} \\
\cline{1-10} \\
\multicolumn{10}{c}{GD 165}} 
\startdata
F1 &8305.96 &120.39543 &\nodata &1.76 &\nodata &-2.73 &1? &-1 &1 \\
\nodata &8308.69 &120.35585 &\nodata &4.79 &\nodata &\nodata &1? &0 &1 \\
\nodata &8311.24 &120.31905 &\nodata &1.36 &\nodata &2.55 &1? &+1 &1 \\
F2 &5187.02 &192.78879 &\nodata &0.85 &\nodata &-2.98 &1 or 2 &-1 &1 \\
\nodata &5190.00 &192.67841 &\nodata &2.35 &\nodata &\nodata &1 or 2 &0 &1 \\
\nodata &5192.82 &192.57373 &\nodata &1.91 &\nodata &2.82 &1 or 2 &+1 &1 \\
F3 \tablenotemark{c} &3989.2 &250.6797 &\nodata &0.6 &\nodata &-7.8 &1 or 2 &-1? &1 \\
\nodata &3997.0 &250.1864 &\nodata &1.0 &\nodata &\nodata &1 or 2 &0 &1 \\
\cutinhead{R548 \tablenotemark{d}}
F1 &4691.915 &213.1326 &\nodata &6.7 &\nodata &\nodata &1 &-1 &2 \\
\nodata &4699.946 &212.7684 &\nodata &4.1 &\nodata &8.031 &1 &+1 &2 \\
F2 &3639.348 &274.7745 &\nodata &2.9 &\nodata &\nodata &1 &-1 &2 \\
\nodata &3646.297 &274.2508 &\nodata &4.1 &\nodata &6.949 &1 &+1 &2 \\
F3 &2997.25 &333.639 &0.001 &1.03 &0.13 &\nodata &1 or 2 &0? &2 \\
F4 &3143.92 &318.074 &0.001 &1.10 &0.13 &\nodata &1 or 2 &0? &2 \\
F5 &5339.43 &187.286 &0.001 &0.43 &0.12 &\nodata &1 or 2 &0? &2 \\
\cutinhead{G226-29}
F1 &9134.7234 &109.47239 &0.00019 &2.82 &0.10 &-16.2175 &1 &-1 &3 \\
\nodata &9150.9409 &109.27838 &0.00051 &1.08 &0.10 &\nodata &1 &0 &3 \\
\nodata &9167.0187 &109.08672 &0.00022 &2.49 &0.10 &16.0778 &1 &+1 &3 \\
\enddata
\tablecomments{ The $\ell$ identifications are from our analysis.}
\tablenotetext{a}{$\Delta f$ is the separation between the modes in the multiplets and the $m=0$ member.}
\tablenotetext{b}{The $m$ identifications for GD 165 are based on frequency splitting alone, not on the size of the combination peak limits.}
\tablenotetext{c}{\citet{ber93} listed F3 from the WET data set, but not from the combined CFHT and WET data set.  They identified the $m=-1$ member as questionable.}
\tablenotetext{d}{The pulsation modes F3, F4, and F5 are those found in the 2001 data set of \citet{muk03}.}
\tablerefs{(1) \citet{ber93}; (2) \citet{muk03}; (3) \citet{kep95a}.}
\end{deluxetable}
\clearpage
\citet{ber93} did not report finding any combination frequencies in the WET and CFHT combined data for GD 165.  In Figure~\ref{RcGD165}, we have plotted the theoretical predictions for GD 165 ($R_c$ with $\Theta_\circ=25^\circ$) along with the observed noise limits for putative combination peaks.  Once again the limits alone are sufficient to constrain F1 tentatively as $\ell=1$, and the other modes to be $\ell \leq 2$, consistent with the $\ell=1$ identifications for each mode determined by \citet{brad01} based on the multiplet structure.  We list our mode identifications in Table~\ref{tbl-pubnocomb}.
\clearpage
\clearpage
\subsubsubsection{\it{R548}}

R548, also called ZZ Ceti, is the prototype of this class of stars, and is one of the brightest and hottest.  \citet{lh71} discovered R548 to be variable.  Its primary pulsation modes are doublets at 213 and 274 seconds.  Like G117-B15A, R548 has a very stable pulsation mode at 213 seconds \citep{muk03}.

The data for R548 were gathered in the WET campaigns XCov 18 in 1999 November and XCov 20 in 2000 November, with additional data from the CFHT and the McDonald Observatory \citep{muk03}.  The pulsation periods and amplitudes for R548 have been taken from \citet{muk03} and are listed in Table~\ref{tbl-pubnocomb}.  As with our assumption for GD 244, \citet{brad98} suggests that the doublet structure in F1 and F2 results from viewing $\ell = 1$ modes at high inclination so that the third (central, $m=0$) mode does not appear.  Using the maximum amplitude of the prewhitened FT of F1, at 213 s, as an estimate on the amplitude of the $m = 0$ peak, we find the minimum possible inclination for R548 to be $79$ degrees.  Unlike GD 244, another high inclination star, we do not detect combination frequencies in R548.  

\begin{figure}
\includegraphics[scale=.7,angle=-90]{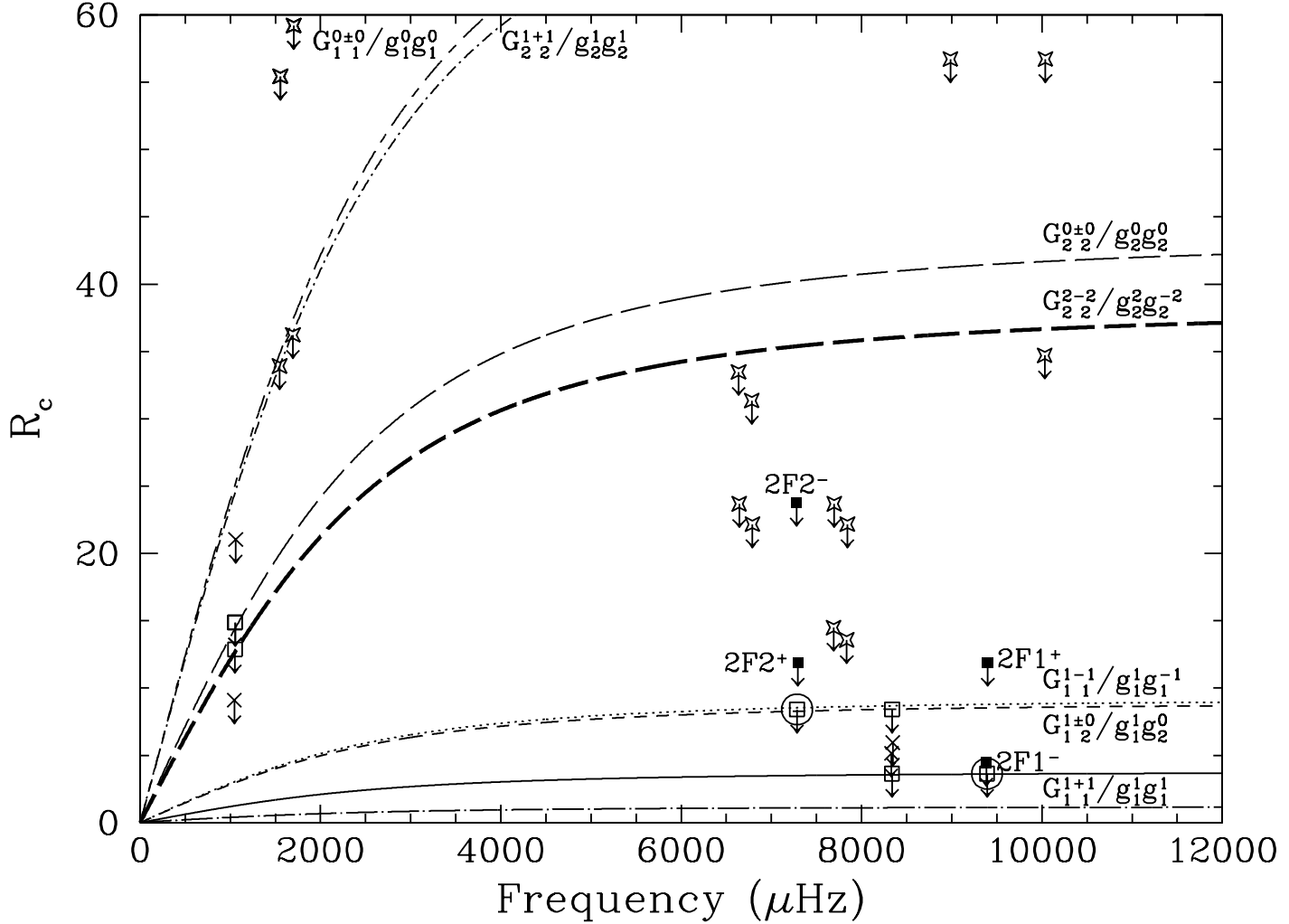}
\caption{Ratio of combination to parent mode amplitudes ($R_c$) for R548.  The lines are theoretical predictions for $G_{1~1}^{0\pm0} \diagup g_1^0 g_1^0 (\Theta_\circ=79^\circ)$ (long-short-dashed line), $G_{2~2}^{1+1} \diagup g_2^1 g_2^1 (\Theta_\circ=79^\circ)$ (dot-dashed line), $G_{2~2}^{0\pm0} \diagup g_2^0 g_2^0 (\Theta_\circ=79^\circ)$ (long-dashed line), $G_{2~2}^{2-2} \diagup g_2^2 g_2^{-2} (\Theta_\circ=79^\circ)$ (bold long-dashed line), $G_{1~1}^{1-1} \diagup g_1^1 g_1^{-1} (\Theta_\circ=79^\circ)$ (dotted line), $G_{1~2}^{1\pm0} \diagup g_1^1 g_2^0 (\Theta_\circ=79^\circ)$ (dashed line), $G_{1~1}^{1+1} \diagup g_1^1 g_1^1 (\Theta_\circ=79^\circ)$ (solid line), $G_{1~1}^{1\pm0} \diagup g_1^1 g_1^0 (\Theta_\circ=79^\circ)$ (solid line, not labeled because of space), and $G_{2~2}^{2+2} \diagup g_2^2 g_2^2 (\Theta_\circ=79^\circ)$ (dot-long-dashed line, not labeled).  The data points are the limits on the harmonics (filled squares), limits for the same-$m$ cross combinations (crosses), limits for the different-$m$ cross combinations (open squares), and limits for the cross combinations between one of the doublet modes (F1 or F2) and one of the lower amplitude singlet modes (stars).  The downward arrows on the limits indicate that the points represent maximum values.  The open square limits for F1$^-$+F1$^+$ and F2$^-$+F2$^+$ are circled for easier identification. }
\label{RcR548}
\end{figure}

In Figure~\ref{RcR548}, we have plotted the predictions for R548 ($R_c$ with $\Theta_\circ=79^\circ$).  We include the observed noise limit (indicated by the filled squares, crosses, open squares, and stars) at the frequencies where we expect combinations to be detected.  The limits on the harmonics of F1 and F2 do not uniquely identify these modes, but only require that $\ell \leq 2$ for both.  2F1$^-$ appears to constrain F1 to be $\ell=1$, but the line below it at the bottom of the Figure is an $\ell=2,2$ combination for $m=2$.  However, the limits on the cross combination peaks at the sum of the $m=-1$ and $m=+1$ parent modes do require F1 and F2 to be $\ell=1$ modes (see the circles in Figure~\ref{RcR548}), which agrees with the identifications of \citet{brad98}.  The limits for the harmonics of the F3, F4, and F5 singlets are sufficient to imply that they are $\ell \leq 2$ if $m=0$, again consistent with the $\ell=2$ identification of \citet{brad98}.  As with L19-2, Wu's predictions suggest that these data are on the threshold of detecting the combination frequencies for the two dominant modes in R548.
\clearpage
\clearpage
\subsubsubsection{\it{G226-29}}

G226-29 is the brightest known ZZ Ceti star because of its proximity \citep[$d=11.0$ pc, $m_v = 12.22$;][]{kep95a}.  J. T. McGraw \& G. Fontaine (1980, unpublished results) discovered its variability, finding only one pulsation mode at 109 s that is rotationally split into a triplet.  G226-29 is the hottest of the 103 known ZZ Ceti stars \citep[see][and references therein]{kep05}, and \citet{kep00b} suggest that we are observing G226-29 just as it enters the instability strip.

\begin{figure} 
\includegraphics[scale=.7,angle=-90]{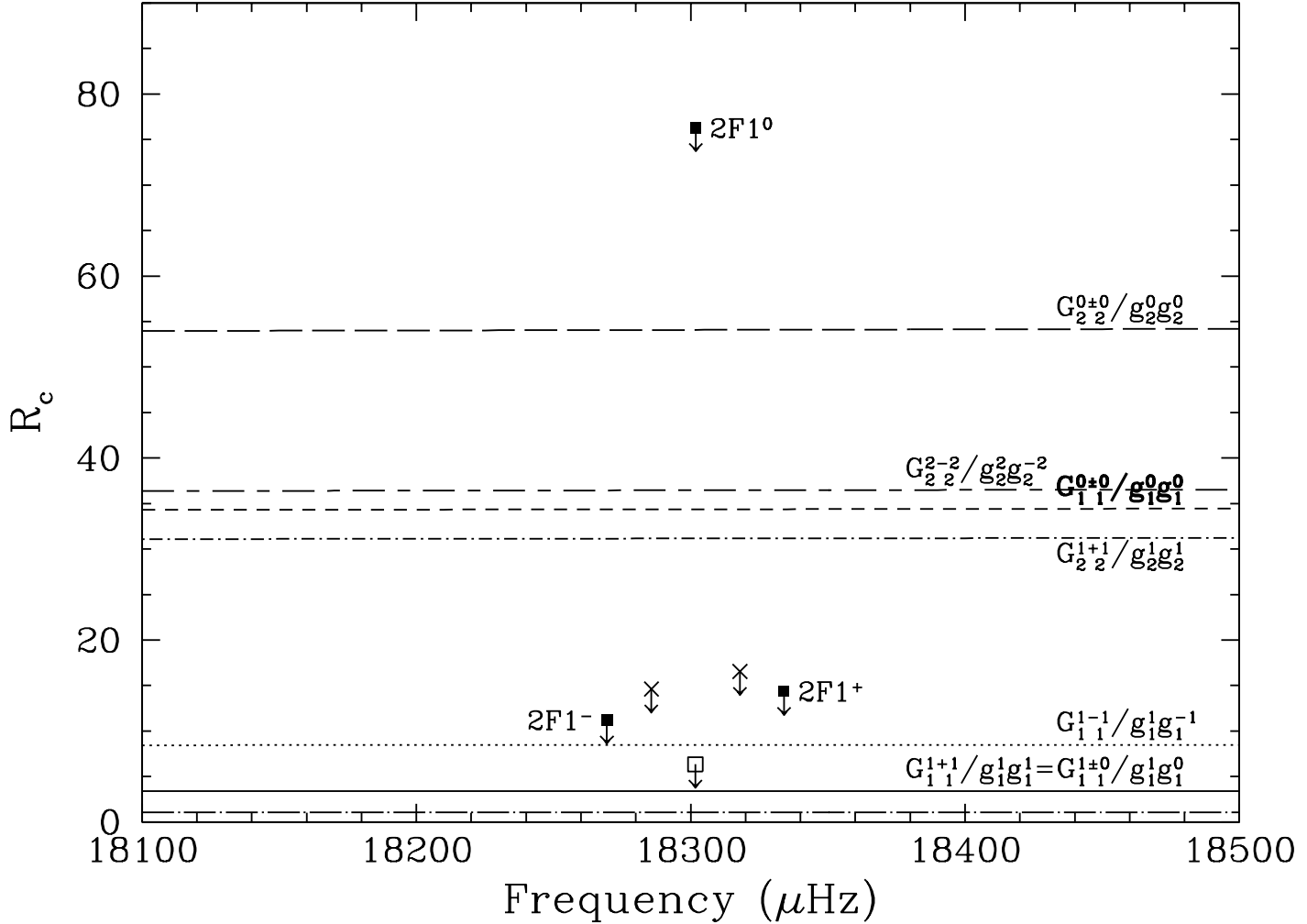}
\caption{Ratio of combination to parent mode amplitudes ($R_c$) for G226-29.  The lines are theoretical predictions for $G_{2~2}^{0\pm0} \diagup g_2^0 g_2^0 (\Theta_\circ=74^\circ)$ (long-dashed line), $G_{2~2}^{2-2} \diagup g_2^2 g_2^{-2} (\Theta_\circ=74^\circ)$ (long-short-dashed line), $G_{1~1}^{0\pm0} \diagup g_1^0 g_1^0 (\Theta_\circ=74^\circ)$ (dashed line), $G_{2~2}^{1+1} \diagup g_2^1 g_2^1 (\Theta_\circ=74^\circ)$ (dot-dashed line), $G_{1~1}^{1-1} \diagup g_1^1 g_1^{-1} (\Theta_\circ=74^\circ)$ (dotted line), $G_{1~1}^{1+1} \diagup g_1^1 g_1^1 (\Theta_\circ=74^\circ)$ (solid line), $G_{1~1}^{1\pm0} \diagup g_1^1 g_1^0 (\Theta_\circ=74^\circ)$ (solid line), and $G_{2~2}^{2+2} \diagup g_2^2 g_2^2 (\Theta_\circ=74^\circ)$ (dot-long-dashed line, not labeled because of space).  Theoretical lines for $G_{2~2}^{2\pm0} \diagup g_2^2 g_2^0 (\Theta_\circ=74^\circ)$ and $G_{2~2}^{1\pm0} \diagup g_2^1 g_2^0 (\Theta_\circ=74^\circ)$ are not shown because they are very near to the predictions for $G_{1~1}^{1\pm0} \diagup g_1^1 g_1^0 (\Theta_\circ=74^\circ)$.  The data points are the limits on the harmonics (filled squares), a limit for the nonzero different-$m$ cross combination (open square), and limits for the cross combinations between the $m\neq 0$ modes and the central $m=0$ mode (crosses).  The downward arrows on the limits indicate that the points represent maximum values.} 
\label{RcG226}
\end{figure}

We included data for G226-29 from the WET campaign XCov 7 in 1992 February presented in \citet{kep95a}.  We found the inclination to be 74 degrees by assuming the evenly spaced triplet is an $\ell=1$ mode.  As discussed previously, our results do not depend sensitively on this assumption.  \citet{kep95a} did not detect combination frequencies in G226-29.  In Figure~\ref{RcG226}, we have plotted the predictions for G226-29 ($R_c$ with $\Theta_\circ=74^\circ$).  We include the observed noise limit (indicated by the filled squares, crosses, and a open squares) at the frequencies where we expect combinations to be detected.  The limits for the harmonics of each member of the triplet do not uniquely identify F1, but only require that $\ell \leq 2$.  As with R548, another high inclination star with detected $m=\pm1$ multiplet members, it is the limits on the cross combination peak at the sum of the $m=-1$ and $m=+1$ parent modes that require the G226-29 F1 mode to be $\ell=1$.  This result is consistent with the time-resolved UV spectroscopy data from HST presented in \citet{kep00b}.  As with L19-2 and R548, Wu's predictions suggest that these data are on the threshold of detecting the combination frequencies in G226-29.
\clearpage
\clearpage
\section{SUMMARY AND CONCLUSIONS}
\label{sum}

The main result of our study is that combination frequencies, particularly harmonics, in the lightcurves of hot ZZ Ceti stars can be used along with the theory of \citet{wu} to constrain and in many cases to determine uniquely the spherical harmonic index ($\ell$) of the modes that produced them.  The first and easiest result to achieve with this method is to identify those modes with $\ell>2$ and those with $\ell \leq 2$.  This alone is useful for significantly reducing the number of seismological models that need to be considered for a given star \citep{brad96}.  The theoretical mode spectrum at $\ell=3$ and higher is so dense that there are many possible model fits to the typically sparse number of detected modes.  By eliminating from consideration the high $\ell$ modes, the possibility of identifying a unique fit is improved.  With the exception of a few small amplitude modes, in this paper we have successfully eliminated $\ell > 2$ identification for all modes in seven of the eight stars in our study.  The eighth star, G185-32, was previously thought to have a high $\ell$ mode \citep{thomp04}, and our method confirms this result (though we get $\ell=3$ instead of $\ell=4$).

For modes with sufficiently large amplitude, combination frequency amplitudes are further able to discriminate between $\ell=1$ and $\ell=2$, primarily through the use of harmonics.  The harmonics are superior for this purpose because they are known to be same-$\ell$ combinations, and because same-$\ell$ combinations are well-separated in the theoretical plots of $R_c$.  We were able to identify modes as $\ell=1$ in six of the eight stars, and in every case our identifications agreed with any previous independent results.

The method we have used requires only time-series photometry and simple calculations as presented in \S\ref{theory}.  The essential part of these calculations is the evaluation of the geometric term in Wu's theory, which we have named ${\cal G}$.  Calculating ${\cal G}$ requires the evaluation of integrals of spherical harmonics in the presence of a limb darkening law.  To assist others in application of this technique, we have included tabulated matrices of combination frequency integrals for $\ell \leq 4$ in Appendix A.  Applying these requires a straightforward estimation of the inclination, which we have done using multiplet amplitudes, where detected, and limits on the sizes of multiplet members where not detected.  This has required that we assume that modes of every $m$ are excited to the same amplitude in every mode, and that rotation always removes the frequency degeneracy of multiplet members.  Fortunately, our results are not highly sensitive to these assumptions.

For convenience, we summarize application of the method as follows:

\noindent 1. Calculate the inclination with the \citet{pes85} method using the ratio of the observed amplitudes in a given multiplet.  Consult the sensitivity of ${\cal G}$ to inclination (see Figures~\ref{Ggg}, \ref{G1111gg}, and \ref{G34}) to ensure that $R_c$ is changing slowly with inclination near this value.

\noindent 2. Calculate the theoretical $R_c$ for both $\ell=1$ and 2 by approximating $\tau_{c_\circ}$ with the longest period mode and using the bolometric correction $\alpha_\lambda=0.4$.  We use $2\beta+\gamma=-9.35$ for hot, low-amplitude DA stars, while Wu uses $2\beta+\gamma=-10$ for the cool DA star G29-38.

\noindent 3. Compare the calculated $R_c$ with the measured value obtained with the amplitudes of the combinations and their parents.

In addition to our application of this method to eight stars, we have presented analyses of new data on GD 66 and GD 244 that will be useful for seismological study of these objects.  We have not been able to definitively decompose the multiplet structure in these stars with single-site data, but the mode periods we have measured are given in Tables~\ref{tbl-GD66} and \ref{tbl-GD244} for comparison to seismological models.

More important than the results for these individual stars is our verification of a quick and easy diagnostic tool that frequently yields definitive results.  We hope the method will find broad and immediate application in the study of numerous ZZ Ceti stars being discovered with the Sloan Digital Sky Survey \citep[see][]{muk04,mul05}.  Most of these are fainter than the objects we have measured, but photometry on a 4-m class telescope will be sufficient to reach useful detection limits.  For example, if the V=15.56 mag star GD 66 were instead an 18th magnitude star, observations with a 4-m class telescope would reveal four of the combination frequency peaks that we identified, including the harmonics of the two highest amplitude peaks.  Further study of the objects in this paper will also be useful, both to secure definitive $\ell$ identifications of the smaller amplitude modes and to detect those combinations that hover just below the detection limits of the present data.  Observations are currently in progress using the 4.1-m SOAR telescope.

\acknowledgments

This work was supported by a CAREER grant from the National Science Foundation (AST 000-94289) and by two grants from the North Carolina Space Grant Consortium.  We thank S. O. Kepler for his helpful comments and suggestions.  We also thank the referee, P. A. Bradley, for reading the paper extremely carefully and offering many suggestions that have improved this paper.

\appendix

\section{SELECTED SOLUTIONS FOR $ G_{\ell_i~\ell_j}^{m_i \pm m_j} \diagup g_{\ell_i}^{m_i} g_{\ell_j}^{m_j}(\Theta_\circ)$}

\begin{figure}
\epsscale{.8}
\plotone{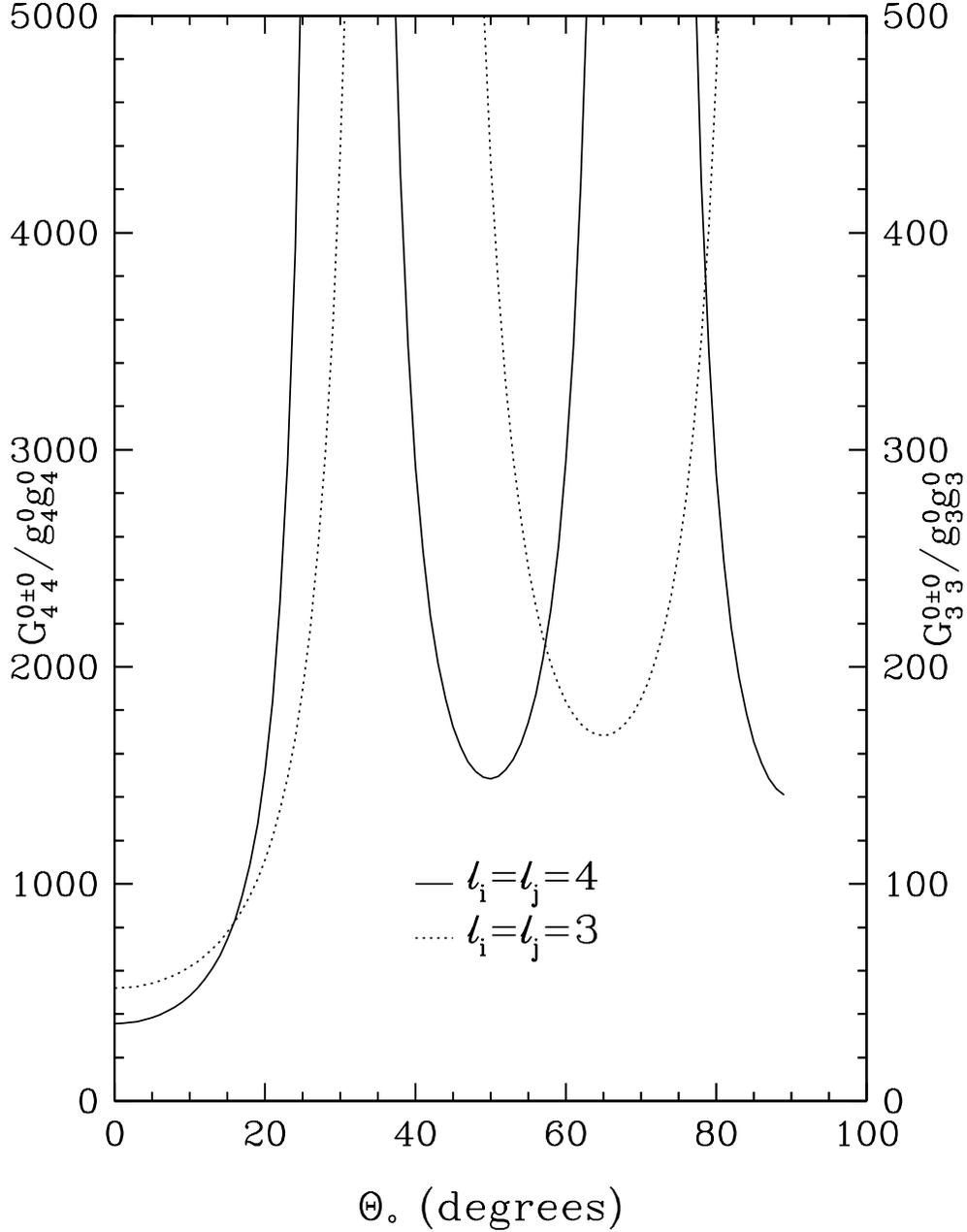}
\caption{${\cal G}$ with $m_i=m_j=0$ (see equation~\ref{Rc}) plotted as a function of inclination angle ($\Theta_\circ$).  The scale of ${\cal G}$ for $\ell_i=\ell_j=4$ (solid line) is a factor of ten larger than the scale of ${\cal G}$ for $\ell_i=\ell_j=3$ (dotted line).  The predicted amplitudes of the combination frequencies show a dramatic increase for inclinations greater than $20^\circ$, and the predictions less than $20^\circ$ increase more rapidly than the cases for $\ell=1$ and 2 (see Figures~\ref{Ggg} and ~\ref{G1111gg}). }
\label{G34}
\epsscale{1.0}
\end{figure}

The following tables contain solutions for $G_{\ell_i~\ell_j}^{m_i \pm m_j} \diagup g_{\ell_i}^{m_i} g_{\ell_j}^{m_j}(\Theta_\circ)$ (see equations~\ref{glm} and ~\ref{Glm}) for the values of $\ell$ and $m$ that are potentially useful for mode identification with photometry using the theory of Wu.  Table~\ref{tbl-G11} contains solutions for $\ell_i=\ell_j=1$.  Tables~\ref{tbl-G12} and \ref{tbl-G22} contain solutions for $\ell_i=1$, $\ell_j=2$ and $\ell_i=\ell_j=2$.  Finally, Table~\ref{tbl-G34} contains solutions for $\ell_i=3$ or 4 and $\ell_j\leq4$ with $m_i=m_j=0$.  For comparison with Figures~\ref{Ggg} and \ref{G1111gg}, we include a plot of the variations of ${\cal G}$ with inclination for $\ell=3,3$ and $\ell=4,4$ in Figure~\ref{G34}.
\clearpage
\begin{deluxetable}{cccc}
\tablecolumns{4}
\tabletypesize{\small}
\tablewidth{0pt}
\tablecaption{Values of $G_{1~1}^{m_i+m_j}\diagup g_1^{m_i}g_1^{m_j}(\Theta_\circ)$  \label{tbl-G11}}
\tablehead{
\colhead{$m_i \backslash m_j$} & \colhead{-1} & \colhead{0} & \colhead{+1}} 
\startdata
-1 &0.65 &0.65 &$-0.65-{{0.90}\over{\sin^2(\Theta_\circ)}}$ \\
\\[5pt]

0 &0.65 &$0.65+{{0.45}\over{\cos^2(\Theta_\circ)}}$ &-0.65 \\
\\[5pt]
+1 &$-0.65-{{0.90}\over{\sin^2(\Theta_\circ)}}$ &-0.65 &0.65 \\
\enddata
\tablecomments{ For values of $G_{1~1}^{m_i-m_j}\diagup g_1^{m_i}g_1^{m_j}(\Theta_\circ)$, reverse the sign of $m_j$. }
\end{deluxetable}


\begin{deluxetable}{cccccc}
\tablecolumns{6}
\tabletypesize{\small}
\tablewidth{0pc}
\tablecaption{Values of $G_{1~2}^{m_i+m_j}\diagup g_1^{m_i}g_2^{m_j}(\Theta_\circ)$  \label{tbl-G12}}
\tablehead{
\colhead{$m_i \backslash m_j$} & \colhead{-2} & \colhead{-1} & \colhead{0} & \colhead{+1} & \colhead{+2}} 
\startdata
-1 &0.27 &0.27 &${{-0.81\sin^2(\Theta_\circ)-0.58}\over{3\cos^2(\Theta_\circ)-1}}$ &$-0.27-{{1.12}\over{\sin^2(\Theta_\circ)}}$ &$0.27+{{2.24}\over{\sin^2(\Theta_\circ)}}$ \\
\\[5pt]
0 &0.27 &$0.27+{{0.56}\over{\cos^2(\Theta_\circ)}}$ &${{0.81\cos^2(\Theta_\circ)+1.97}\over{3\cos^2(\Theta_\circ)-1}}$ &$-0.27-{{0.56}\over{\cos^2(\Theta_\circ)}}$ &0.27 \\
\\[5pt]
+1 &$-0.27-{{2.24}\over{\sin^2(\Theta_\circ)}}$ &$-0.27-{{1.12}\over{\sin^2(\Theta_\circ)}}$ &${{0.81\sin^2(\Theta_\circ)+0.58}\over{3\cos^2(\Theta_\circ)-1}}$ &0.27 &-0.27 \\
\enddata
\tablecomments{ For values of $G_{1~2}^{m_i-m_j}\diagup g_1^{m_i}g_2^{m_j}(\Theta_\circ)$, reverse the sign of $m_j$.}

\end{deluxetable}

\clearpage
\thispagestyle{empty}
\begin{deluxetable}{cccccc}
\tablecolumns{6}
\tabletypesize{\tiny}
\rotate
\tablewidth{0pt}
\tablecaption{Values of $G_{2~2}^{m_i+m_j}\diagup g_2^{m_i}g_2^{m_j}(\Theta_\circ)$  \label{tbl-G22}}

\tablehead{
\colhead{$m_i \backslash m_j$} & \colhead{-2} & \colhead{-1} & \colhead{0} & \colhead{+1} & \colhead{+2}} 
\startdata
-2 &-0.20 &-0.20 &${{0.59\cos^4(\Theta_\circ)+1.08\cos^2(\Theta_\circ)-1.67}\over{\sin^2(\Theta_\circ)(3\cos^2(\Theta_\circ)-1)}}$ &$0.20-{{1.87}\over{\sin^2(\Theta_\circ)}} $ &${{-0.20\sin^4(\Theta_\circ)+3.74\sin^2(\Theta_\circ)+2.66}\over{\sin^4(\Theta_\circ)}}$ \\
\\[5pt]
-1 &-0.20 &${{0.20\cos^4(\Theta_\circ)-0.66\cos^2(\Theta_\circ)+0.47}\over{\cos^2(\Theta_\circ)\sin^2(\Theta_\circ)}}$ &${{-0.59\cos^2(\Theta_\circ)+1.13}\over{3\cos^2(\Theta_\circ)-1}}$ &$-{{0.20\cos^4(\Theta_\circ)+0.27\cos^2(\Theta_\circ)+1.13}\over{\cos^2(\Theta_\circ)\sin^2(\Theta_\circ)}}$ &$-0.20+{{1.87}\over{\sin^2(\Theta_\circ)}}$ \\
\\[5pt]
0 &${{0.59\cos^4(\Theta_\circ)+1.08\cos^2(\Theta_\circ)-1.67}\over{\sin^2(\Theta_\circ)(3\cos^2(\Theta_\circ)-1)}}$ &${{-0.59\cos^2(\Theta_\circ)+1.13}\over{3\cos^2(\Theta_\circ)-1}}$ &${{-1.78\cos^4(\Theta_\circ)+6.80\cos^2(\Theta_\circ)+5.66}\over{(3\cos^2(\Theta_\circ)-1)^2}}$ &${{0.59\cos^2(\Theta_\circ)-1.13}\over{3\cos^2(\Theta_\circ)-1}}$ &${{0.59\cos^4(\Theta_\circ)+1.08\cos^2(\Theta_\circ)-1.67}\over{\sin^2(\Theta_\circ)(3\cos^2(\Theta_\circ)-1)}}$ \\
\\[5pt]
+1 &$0.20-{{1.87}\over{\sin^2(\Theta_\circ)}}$ &$-{{0.20\cos^4(\Theta_\circ)+0.27\cos^2(\Theta_\circ)+1.13}\over{\cos^2(\Theta_\circ)\sin^2(\Theta_\circ)}}$ &${{0.59\cos^2(\Theta_\circ)-1.13}\over{3\cos^2(\Theta_\circ)-1}}$ &${{0.20\cos^4(\Theta_\circ)-0.66\cos^2(\Theta_\circ)+0.47}\over{\cos^2(\Theta_\circ)\sin^2(\Theta_\circ)}}$ &0.20 \\
\\[5pt]
+2 &${{-0.20\sin^4(\Theta_\circ)+3.74\sin^2(\Theta_\circ)+ 2.66}\over{\sin^4(\Theta_\circ)}}$ &$-0.20+{{1.87}\over{\sin^2(\Theta_\circ)}} $ &${{0.59\cos^4(\Theta_\circ)+1.08\cos^2(\Theta_\circ)-1.67}\over{\sin^2(\Theta_\circ)(3\cos^2(\Theta_\circ)-1)}}$ &0.20 &-0.20 \\

\enddata
\tablecomments{ For values of $G_{2~2}^{m_i-m_j}\diagup g_2^{m_i}g_2^{m_j}(\Theta_\circ)$, reverse the sign of $m_j$.}
\end{deluxetable}

\begin{deluxetable}{ccc}
\tablecolumns{3}
\tabletypesize{\small}
\tablewidth{0pt}
\tablecaption{Values of $G_{\ell_i~\ell_j}^{0\pm 0}\diagup g_{\ell_i}^0 g_{\ell_j}^0(\Theta_\circ)$  \label{tbl-G34}}
\tablehead{
\colhead{$\ell_i \backslash \ell_j$} & \colhead{3} & \colhead{4}} 

\startdata
1 &${{-0.78\cos^4(\Theta_\circ)+3.82\cos^2(\Theta_\circ)-1.12}\over{1.67\cos^4(\Theta_\circ)- \cos^2(\Theta_\circ)}}$ &${{-6.18\cos^4(\Theta_\circ)+19.41\cos^2(\Theta_\circ)-9}\over{-11.67\cos^4(\Theta_\circ)+10\cos^2(\Theta_\circ)-1}}$ \\
\\[5pt]
2 &${{-1.92\cos^4(\Theta_\circ)+4.87\cos^2(\Theta_\circ)+9.86}\over{(3\cos^2(\Theta_\circ)-1)(1.67\cos^2(\Theta_\circ)-1)}}$ &${{40.38\cos^6(\Theta_\circ)-73.72\cos^4(\Theta_\circ)+144.05\cos^2(\Theta_\circ)-39.04}\over{(3\cos^2(\Theta_\circ)-1)(-11.67\cos^4(\Theta_\circ)+10\cos^2(\Theta_\circ)-1)}}$ \\
\\[5pt]
3 &${{5.56\cos^6(\Theta_\circ)-10\cos^4(\Theta_\circ)+15.17\cos^2(\Theta_\circ)+12.40}\over{(1.67\cos^3(\Theta_\circ)-\cos(\Theta_\circ))^2}}$ &${{35\cos^6(\Theta_\circ)-75\cos^4(\Theta_\circ)+85\cos^2(\Theta_\circ)+153.69}\over{(1.67\cos^2(\Theta_\circ)-1)(-11.67\cos^4(\Theta_\circ)+10\cos^2(\Theta_\circ)-1)}}$ \\
\\[5pt]
4 &${{35\cos^6(\Theta_\circ)-75\cos^4(\Theta_\circ)+85\cos^2(\Theta_\circ)+153.69}\over{(1.67\cos^2(\Theta_\circ)-1)(-11.67\cos^4(\Theta_\circ)+10\cos^2(\Theta_\circ)-1)}}$ &${{-1225\cos^8(\Theta_\circ)+2660\cos^6(\Theta_\circ)-2070\cos^4(\Theta_\circ)+1769.26\cos^2(\Theta_\circ)+1400.80}\over{(-11.67\cos^4(\Theta_\circ)+10\cos^2(\Theta_\circ)-1)^2}}$ \\
\enddata
\end{deluxetable}
\clearpage
\clearpage

\end{document}